\documentclass[lettersize,journal]{IEEEtran}
\usepackage{amsmath,amssymb,amsfonts}
\usepackage{algorithmic}
\usepackage{algorithm}
\usepackage{array}
\usepackage[caption=false,font=normalsize,labelfont=sf,textfont=sf]{subfig}
\usepackage{textcomp}
\usepackage{stfloats}
\usepackage{url}
\usepackage{verbatim}
\usepackage{graphicx}
\usepackage{cite}
\usepackage{xcolor} % added for changing color
\usepackage{bm}   % ← added to fix bug of "\bm undefined"
\begin{document}

\title{Continuous-Discrete Dual-Domain Co-Design for Optimized Digital Control of Power Converters}

\author{Shen Chen,~\IEEEmembership{Student Member,~IEEE,} Jinjun Liu,~\IEEEmembership{Fellow,~IEEE,} Xuancai Zhu, Yanlong Li, Jisong Wang, Shuaipo Guo, Zhipeng Song, Lijun Hang,~\IEEEmembership{Senior Member,~IEEE,} Zeng Liu,~\IEEEmembership{Senior Member,~IEEE} 
        % <-this % stops a space
%\thanks{This paper was produced using the \LaTeX template released by the IEEE Publication Technology Group.}% <-this % stops a space
\thanks{ Manuscript received Month xx, 2026; 
This work was supported by the National Key Research and Development Program of China under Grant 2023YFB2604600.
(Corresponding author: Jinjun Liu.)

Shen Chen is with the School of Electrical Engineering, Xi'an Jiaotong University, Xi'an 710049, China,
and also with SolaX Power Network Technology (Zhejiang) Co., Ltd., Hangzhou 310013, China. 
(e-mail: chenshen@stu.xjtu.edu.cn).

Jinjun Liu, and Zeng Liu are with the State Key Laboratory of Electrical Insulation and
Power Equipment, School of Electrical Engineering, Xi'an Jiaotong University, Xi'an 710049, China. 
(e-mail: jjliu@mail.xjtu.edu.cn; zengliu@mail.xjtu.edu.cn).

Xuancai Zhu is with Hoymiles Power Electronics Inc., Hangzhou 310013, China.
(e-mail: rex.zhu@hoymiles.com).

Yanlong Li, Jisong Wang, Shuaipo Guo, and Zhipeng Song are with SolaX Power Network Technology (Zhejiang) Co., Ltd., Hangzhou 310013, China. 
(e-mail: liyanlong@solaxpower.com; wangjisong@solaxpower.com; guoshuaipo@solaxpower.com; songzhipeng@solaxpower.com).

Lijun Hang is with the Regional Energy Internet Technology Zhejiang Engineering
Laboratory, School of Automation, Hangzhou Dianzi University, Hangzhou 310018, China.
(e-mail: ljhang@hdu.edu.cn).
}
}

% The paper headers
\markboth{IEEEtran \LaTeX\ Template v1.8b,~Vol.~x, No.~x, October~2025}%
{Shell \MakeLowercase{\textit{et al.}}: A Sample Article Using IEEEtran.cls for IEEE Journals}

\maketitle

\begin{abstract}
Optimizing digital control for power converters typically involves designing 
continuous-time controllers and discretizing them via Euler, Tustin, or Generalized Bilinear Transform (GBT) methods. 
However, the intrinsic coupling between continuous-domain control parameters and 
discrete-domain implementation parameters remains underexplored, 
often leading to suboptimal performance. 
To bridge this gap, this paper first establishes a Non-Uniqueness Theorem, 
elucidating the coupling mechanism between dual-domain parameters. 
Based on this, two co-design methodologies are proposed to derive optimized parameter sets: 
the Explicit Frequency-Domain Analytical Method (EFDAM) and 
the Implicit Computer-Aided Search Method (ICASM). 
The EFDAM derives closed-form solutions by maximizing the crossover frequency and 
resonant magnitude via logarithmic frequency stability criteria, 
while minimizing discretization errors using Taylor-series bounds. 
The ICASM formulates the search for optimized parameters as a constrained multi-objective optimization problem, 
balancing bandwidth, resonant magnitude, and fundamental magnitude. 
To solve ICASM efficiently, two algorithms are introduced: 
the Multi-Start COBYLA (MS-COBYLA), 
which employs Latin Hypercube Sampling for physics-informed initialization to escape local optima, 
and the PSO-COBYLA hybrid algorithm, 
which couples global particle swarm exploration with local refinement. 
Experimental results on a 5-kW single-phase inverter with 
a Proportional-Integral-Resonant (PIR) regulator validate that 
the proposed methods outperform the conventional benchmark, 
achieving a 43.3\% reduction in the peak-to-peak current ripple measured at the
zero-active-power operating point and 
an 11.7\%--35.9\% decrease in THDi across various harmonic profiles.
\end{abstract}


\section{Introduction}
\IEEEPARstart{C}{urrent} regulation performance significantly impacts the operation of grid-tied inverters, 
demanding optimized design of control parameters and 
discretization strategies \cite{teodorescu_ipepa_2006, holmes_tpel_2009, zamani_tie_2024}. 
To eliminate steady-state errors in AC current tracking, resonant regulators—
such as Proportional-Integral-Resonant (PIR) and Proportional-Resonant (PR) controllers—
have been extensively adopted 
\cite{husev_ietpel_2015, komurcugil_tie_2016, komurcugil_tie_2017, altin_tie_2019, lin_spp_tpel_2022, guler_tie_2024, chen_tie_2024}, 
leveraging the internal model principle to achieve precise sinusoidal tracking.

Recent works address the integration of resonant regulators with LCL filters \cite{bao_tpel_2014}, 
where the intrinsic resonance of the third-order LCL filter poses stability challenges.
Representative solutions include weighted-average current feedback to reduce system order~\cite{shen_tie_2010}, 
capacitor-voltage feedforward to cancel resonance frequencies~\cite{citro_tie_2016}, 
dual-loop structures combining PR with uncertainty/disturbance estimators~\cite{xiong_tpel_2023}, 
passivity-oriented phase-corrected resonant controllers~\cite{geng_tpel_2026a}, 
proportional-derivative-resonant (PDR) controllers with impedance phase shaping~\cite{geng_tpel_2026b}, 
and observer-based full-state feedback for sensorless operation, 
which indirectly mitigates resonance by eliminating the need for sensors of grid voltage~\cite{fantino_tie_2019}.
Other studies examine resonant regulators under complex grid conditions, 
such as decentralized harmonic mitigation against background distortions~\cite{jafarian_jestpe_2018}, 
impedance-model-based stability comparisons between PI and PR schemes~\cite{yang_jestpe_2022}, 
and time-domain stability analysis via Floquet theory~\cite{li_tie_2021}.
These investigations collectively reveal that PR-based current loops exhibit both resilience and limitations 
when subjected to grid impedance variation, unbalanced loading, and harmonic pollution. 
Further efforts extend resonant regulators across diverse inverter topologies. 
Single-phase systems include passive-damped LCL/LLCL inverters~\cite{wu_tie_2013}, 
quasi-Z-source inverters with virtual time constants~\cite{komurcugil_tie_2018}, 
and flyback microinverters~\cite{lee_tie_2016}. 
Three-phase and grid-forming applications span Boost inverters~\cite{chen_tpel_2020}, 
multi-level switched-capacitor and T-type inverters~\cite{srivastava_tec_2023, ravali_cpss_2026}, 
transformerless H5 PV inverters~\cite{mahfuz_tie_2023}, 
triple line-voltage cascaded converters~\cite{xia_tie_2013}, 
three-phase grid-forming inverters~\cite{quan_tie_2021}, 
and SPV-DSTATCOMs under partial shading~\cite{dash_access_2021}. 
Collectively, these works position resonant regulators as a mature yet evolving paradigm in modern grid-tied inverters.

In addition, extensive efforts have been devoted to parameter tuning of resonant regulators. 
Representative works include one-dimensional optimization under 
source impedance variation~\cite{chen_tie_2014}, anti-windup schemes for 
single-phase PV systems~\cite{errouissi_tia_2021}, 
adaptive PR tuning via frequency-locked loops~\cite{seifi_tie_2019}, 
phase-lag exploitation for robust stability under fluctuating grid impedance~\cite{chen_tie_2025}, 
per-phase PR offset suppression for split DC-link topologies~\cite{fadlika_tie_2026}, 
and resonant current controllers with enhanced transient response~\cite{busada_tie_2018}.
In a closely related effort,
\cite{pan_tpel_2017} proposes a unified model for LCL-type inverter current control, 
simplifying parameter tuning. 
However, the method still uses heuristic coefficients; 
e.g., the maximum crossover frequency is set to \(4\%\) of the sampling frequency, 
which is not rigorously optimal.

Nevertheless, the transition to digital control, while offering greater flexibility, 
introduces pronounced sensitivity to discretization errors,
necessitating judicious discretization method selection and parameter optimization \cite{chen_gbt_tii_2026}. 
It is well established that the discretization rule alters both the pole/zero locations and the effective 
frequency scale of the regulator: Euler and Tustin mappings introduce frequency warping that 
displaces the resonant peak away from $\omega_n$, so the gain required for zero 
steady-state error is no longer obtained at the intended frequency \cite{yepes_tpel_2010}; 
the same mechanism also perturbs the controller zeros and therefore the stability margins \cite{yepes_tie_2011}. 
Conventional optimization methods typically follow a sequential workflow: 
control parameters are first optimized in the continuous domain and then discretized 
(most commonly via the prewarped Tustin method). 
Although foundational studies such as \cite{holmes_tpel_2009} emphasize
the importance of incorporating digital delays and distortions into regulator optimization, 
this decoupled approach overlooks the inherent coupling between continuous-domain control parameters and 
discrete-domain implementation parameters, potentially yielding suboptimal performance. 
Recent efforts have attempted to address this limitation by directly optimizing discretization parameters; 
for instance, \cite{chen_gbt_tii_2026} proposes a unified optimization framework that accounts for both magnitude and phase errors. 
However, such optimizations are invariably performed for a given continuous design. 
What has not been clarified is how these discrete-domain parameters interact with the continuous-domain 
parameters, and no choice made inside a single domain can be optimal for the pair. 

Despite these advancements, 
a critical research gap persists: the absence of a systematic framework for the joint optimization of 
continuous-domain control parameters and discrete-domain implementation parameters. 
To bridge this gap, this article proposes a dual-domain co-design methodology. 
We first elucidate the coupling mechanism and formally prove a \textit{non-uniqueness theorem}, 
demonstrating that traditional single-domain optimizations yield only conditional optima. 
Based on this insight, we propose two distinct methodologies:
(1) \textbf{Explicit Frequency-Domain Analytical Method (EFDAM)}: 
Derives closed-form solutions via Bode plot approximations for rapid deployment, but at the expense of suboptimal performance.
(2) \textbf{Implicit Computer-Aided Search Method (ICASM)}: 
Formulates tuning as a constrained black-box optimization problem, 
employing derivative-free algorithms (MS-COBYLA and PSO-COBYLA) to explore the joint space for 
high-quality solutions that are inaccessible in closed form. 
Since the resulting problem is non-convex and the solvers are derivative-free and stochastic, 
the solutions reported in this article are to be understood as best-found (empirically near-global) 
optima rather than certified global optima; this terminology is used consistently hereafter.

The efficacy of joint optimization is validated using a digital PIR controller on a 5-kW grid-tied inverter, 
outperforming conventional two-step designs that typically fall short of theoretical performance limits. 
The remainder of this article is organized as follows: 
Section~\ref{sec2} details system modeling and formalizes the joint optimization problem. 
Section~\ref{sec3} presents the EFDAM with analytical derivations. 
Section~\ref{sec4} elaborates on the ICASM framework, encompassing objective construction, 
constraint handling, and optimization algorithm design. 
Section~\ref{sec5} provides comprehensive experimental validation across steady-state and dynamic scenarios. 
Finally, Section~\ref{sec6} concludes the article.

%-------------------------------------------------------------------------------
% Section II: System Modeling and Problem Formulation
%-------------------------------------------------------------------------------
\section{System Modeling and Problem Formulation}\label{sec2}
\subsection{Continuous-Discrete Dual-Domain Model}
\begin{figure}[htbp] % h,t,b: here, top, bottom
  \centering
  \includegraphics[width=1.0\linewidth]{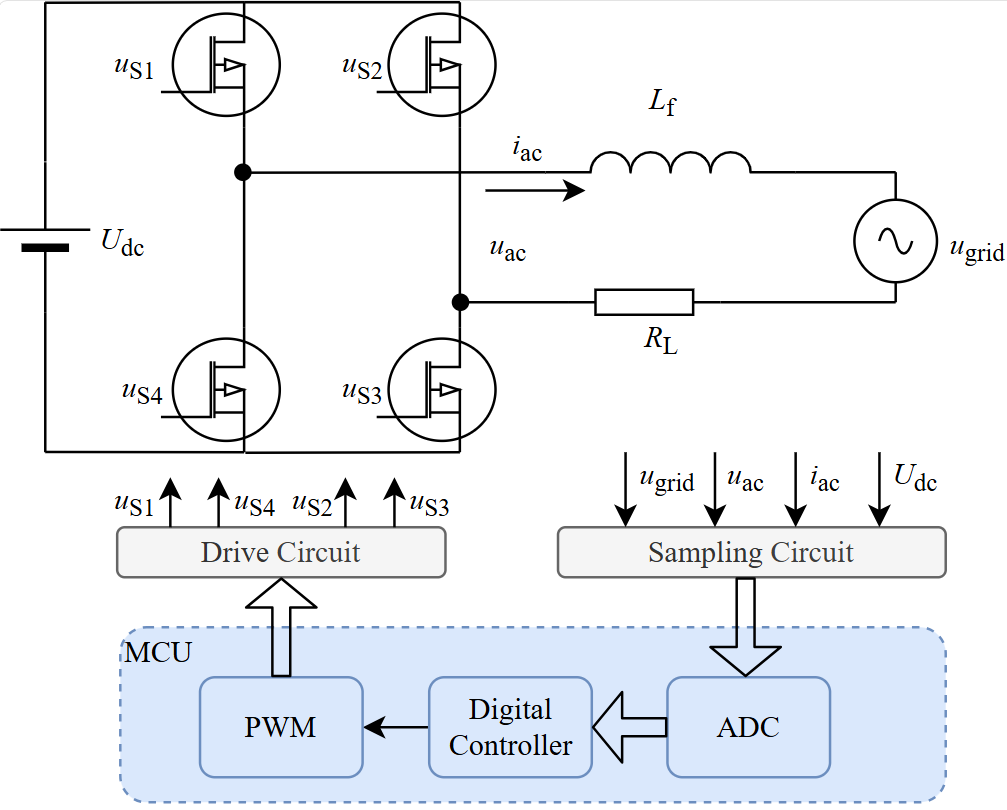}
  \caption{System diagram of a single-phase grid-tied inverter.}
  \label{fig:system_diagram}
\end{figure}

\begin{figure}[htbp] % h,t,b: here, top, bottom
  \centering
  \includegraphics[width=1.0\linewidth]{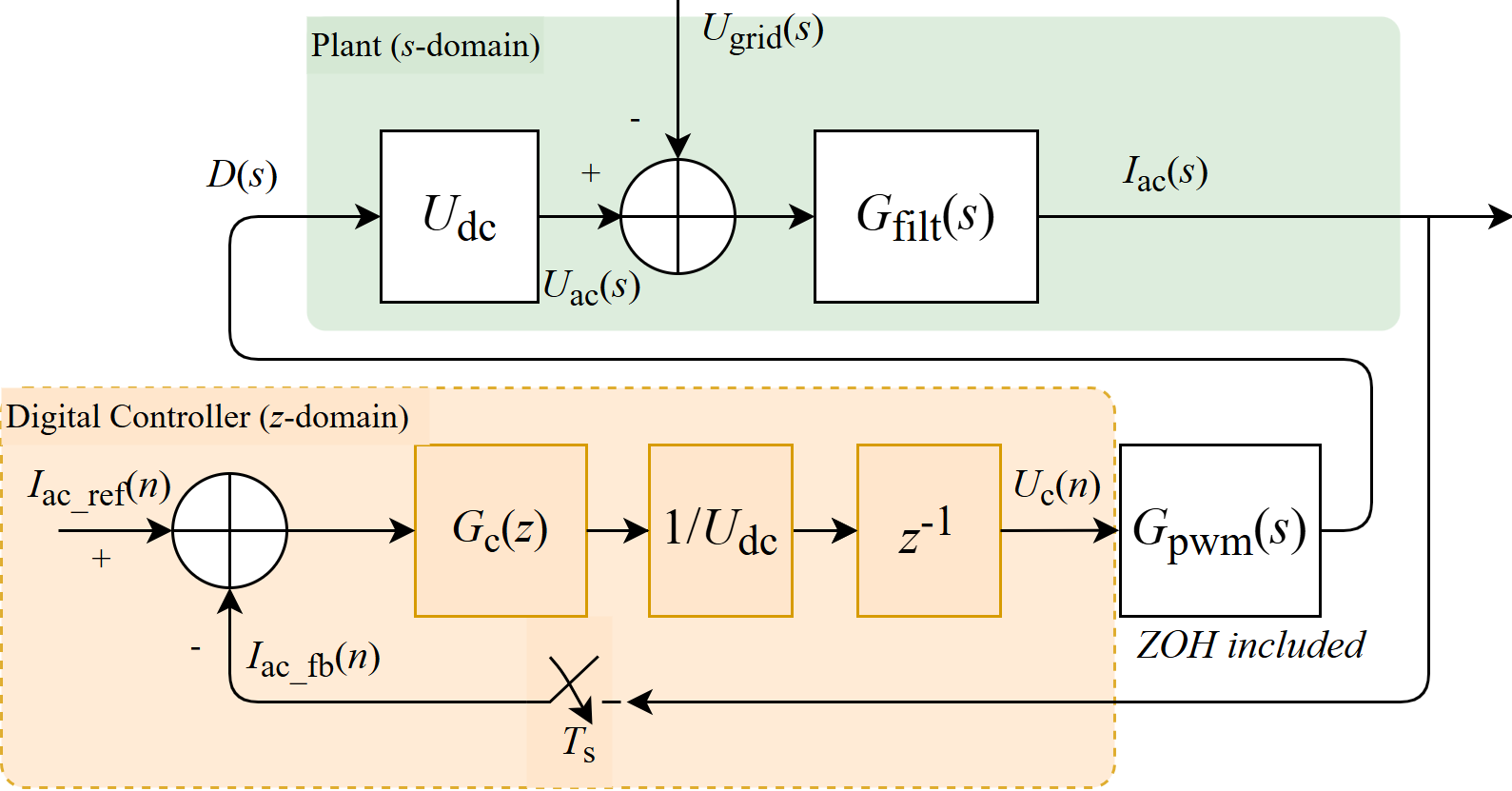}
  \caption{Continuous-discrete dual-domain model of a single-phase grid-tied inverter, including control and PWM delays.}
  \label{fig:dual_domain_model}
\end{figure}

Fig.~\ref{fig:system_diagram} illustrates the system diagram of a single-phase grid-tied inverter. 
The core of the Microcontroller Unit (MCU) comprises Analog-to-Digital Conversion (ADC), 
Pulse Width Modulation (PWM), and the execution of digital control algorithms, 
specifically the PIR controller adopted in this work.
Correspondingly, 
Fig.~\ref{fig:dual_domain_model} depicts the continuous-discrete dual-domain model incorporating control and PWM delays. 
In this model, $I_{\mathrm{ac\_ref}}(n)$ denotes the discrete-time sampled value of the reference current. 
Likewise, $I_{\mathrm{ac\_fb}}(n)$ and $U_{\mathrm{c}}(n)$ represent the sampled feedback current and the controller output voltage, respectively.

The transfer functions of the plant components are defined as follows.
The filter $G_{\mathrm{filt}}(s)$ is modeled as:
\begin{equation}
    G_{\mathrm{filt}}(s) = \frac{1}{s L_{\mathrm{f}} + R_{\mathrm{L}}}, \label{eq:cont_filter}
\end{equation}
where $L_{\mathrm{f}}$ denotes the filter inductance and $R_{\mathrm{L}}$ represents its equivalent series resistance.
Since the crossover frequency $f_\mathrm{c}$ is far below the LCL resonance frequency, 
the filter capacitor's influence on the loop gain is negligible at $f \le f_\mathrm{c}$ \cite{bao_tpel_2014}, 
making the LCL plant effectively equivalent to a single inductor $L_1+L_2$ within the control bandwidth.
Throughout this article $L_\mathrm{f}$ denotes the equivalent inductance seen by the controller within the
control bandwidth: $L_\mathrm{f}=L_1+L_2$ for an LCL filter and $L_\mathrm{f}=L$ for a single-inductor (L)
filter, which is the case realized in the prototype of Section~\ref{sec5} and listed in
Table~\ref{tab:test_system_params}. Since the LCL resonance lies well above $f_\mathrm{c}$, it is not
actively damped and does not enter the loop-shaping constraints.

The PWM module transfer function $G_{\mathrm{pwm}}(s)$ is modeled as:
\begin{equation}
G_{\mathrm{pwm}}(s) = \frac{1-e^{-s T_{\mathrm{s}}}}{s T_{\mathrm{s}}},\label{eq:cont_pwm}
\end{equation}
which represents the zero-order hold (ZOH) effect of the digital pulse-width modulation process 
with sampling period $T_{\mathrm{s}}$.

The computational delay inherent in digital control is characterized as $G_{\mathrm{delay}}(z)=z^{-1}$, 
which accounts for the one-sample-period latency introduced by the control algorithm execution and update mechanism.

The digital controller $G_{\mathrm{c}}(s) = G_{\mathrm{PIR}}(s)$ is given by:
\begin{equation}
G_{\mathrm{PIR}}(s)
= K_{\mathrm{p}} \left( 1 + \frac{1}{\tau_{\mathrm{i}} s} \right)
+ \frac{2 K_{\mathrm{r}} \omega_{\mathrm{b}} s}
{s^2 + 2 \omega_{\mathrm{b}} s + \omega_n^2},
\label{eq:cont_pir}
\end{equation}
where the continuous-domain control parameters comprise the proportional gain $K_{\mathrm{p}}$, 
the integral time constant $\tau_{\mathrm{i}}$, 
the resonant gain $K_{\mathrm{r}}$, 
the resonant bandwidth coefficient $\omega_{\mathrm{b}}$, 
and the resonant angular frequency $\omega_n$.

The Scalable Bilinear Transformation (SBT) \cite{chen_sbt_tpel_2026} has been established as a unified discretization framework 
based on numerical integration, proving particularly effective for PIR controllers. 
Therefore, it is adopted for the discretization of the PIR controller. 
The mapping from the $s$-domain to the $z$-domain is defined as:
\begin{equation}
    s = f(z) = \frac{1}{\beta T_{\mathrm{s}}}\frac{z-1}{\alpha z + (1-\alpha)}, \label{eq:sbt}
\end{equation}
where $\alpha$ denotes the shape factor, $\beta$ represents the time-scaling factor.

Following SBT discretization, the overall discrete transfer function of the PIR controller 
is derived as the parallel summation of its constituent PI and QR components,
\begin{equation}
    \begin{split}
        G_{\mathrm{c}}(z) &= G_{\mathrm{PIR}}(z) = G_{\mathrm{PI}}(z) + G_{\mathrm{QR}}(z) = \\
        &\frac{A_0 z + A_1}{z - 1} + \frac{C_0 z^2 + C_1 z + C_2}{z^2 + D_1 z + D_2}, \label{eq:disc_pir}
    \end{split}
\end{equation}
where the coefficients for the PI controller are defined as $A_0 = K_{\mathrm{p}} (1 + \frac{\alpha \beta T_{\mathrm{s}}}{\tau_{\mathrm{i}}})$,
and $A_1 = K_{\mathrm{p}} (-1 + \frac{(1-\alpha) \beta T_{\mathrm{s}}}{\tau_{\mathrm{i}}})$.

The coefficients for the QR controller are derived using the auxiliary variables $\Psi_1= \frac{1}{(\beta T_{\mathrm{s}})^2} + \frac{2\omega_{\mathrm{b}}\alpha}{\beta T_{\mathrm{s}}} + \omega_n^2\alpha^2$, 
$\Psi_2 = -\frac{2}{(\beta T_{\mathrm{s}})^2} + \frac{2\omega_{\mathrm{b}}(1-2\alpha)}{\beta T_{\mathrm{s}}} + 2\omega_n^2\alpha(1-\alpha)$, 
$\Psi_3 = \frac{1}{(\beta T_{\mathrm{s}})^2} - \frac{2\omega_{\mathrm{b}}(1-\alpha)}{\beta T_{\mathrm{s}}} + \omega_n^2(1-\alpha)^2$,
and the resulting coefficients are:
\begin{align}
D_1 &= \frac{\Psi_2}{\Psi_1}, \quad D_2 = \frac{\Psi_3}{\Psi_1}, \quad C_0 = \frac{2 K_{\mathrm{r}} \omega_{\mathrm{b}} \alpha}{\beta T_{\mathrm{s}} \, \Psi_1}, \label{eq:qr_coeff_1}\\[6pt]
C_1 &= \frac{2 K_{\mathrm{r}} \omega_{\mathrm{b}} (1-2\alpha)}{\beta T_{\mathrm{s}} \, \Psi_1}, \quad C_2 = -\frac{2 K_{\mathrm{r}} \omega_{\mathrm{b}} (1-\alpha)}{\beta T_{\mathrm{s}} \, \Psi_1}.\label{eq:qr_coeff_2}
\end{align}

Table~\ref{tab:param_map} makes the coupling explicit.
Two observations follow directly and are used throughout the rest of the article.
First, a continuous-domain parameter never appears alone: $\tau_\mathrm{i}$ enters only through the
dimensionless ratio $\beta T_\mathrm{s}/\tau_\mathrm{i}$, $\omega_n$ and $\omega_\mathrm{b}$ enter
through $\Psi_{1}$--$\Psi_{3}$ together with $\alpha$ and $\beta$, and $K_\mathrm{r}$ is always
multiplied by $\omega_\mathrm{b}/\Psi_1$.
Second, $(\alpha,\beta)$ enter both branches, so they cannot be interpreted as a pure
``numerical-integration'' choice that is independent of loop shaping: $\alpha$ also moves the
discrete poles $D_1,D_2$ and $\beta$ also rescales the integrator through $A_0,A_1$.

\begin{table}[htbp]
\centering
\caption{Parameter-to-Coefficient Map and Performance Implications}
\label{tab:param_map}
\begin{tabular}{|p{1.1cm}|p{2.4cm}|p{3.9cm}|}
\hline
Parameter & Enters through & Dominant performance effect \\
\hline
$K_\mathrm{p}$ & common factor of $A_0,A_1$ & loop gain $\rightarrow$ crossover frequency $\omega_\mathrm{c}$ and PM \\
$\tau_\mathrm{i}$ & $\alpha\beta T_\mathrm{s}/\tau_\mathrm{i}$, $(1-\alpha)\beta T_\mathrm{s}/\tau_\mathrm{i}$ & low-frequency gain; integral phase lag $\arctan\!\big(1/(\omega_\mathrm{c}\tau_\mathrm{i})\big)$ at crossover \\
$K_\mathrm{r}$ & common factor of $C_0$--$C_2$ & height of the resonant peak $\rightarrow$ harmonic rejection \\
$\omega_\mathrm{b}$ & $\Psi_1$--$\Psi_3$ ($D_1,D_2$) & width/selectivity of the peak $\rightarrow$ tolerance to grid-frequency drift \\
$\omega_n$ & $\Psi_1$--$\Psi_3$ ($D_1,D_2$) & angular position (pole angle) of the peak \\
$\alpha$ & $\Psi_1$--$\Psi_3$, $A_0,A_1$ & discrete pole placement; second-order mapping error $(\tfrac12-\alpha)$ \\
$\beta$ & $\Psi_1$--$\Psi_3$, $A_0,A_1$ & frequency-axis scaling (warping); first-order error $(1-\beta)$ \\
$T_\mathrm{s}$ & all coefficients & \emph{fixed by hardware}; sets the delay budget $1.5T_\mathrm{s}$ \\
\hline
\end{tabular}
\end{table}

$T_\mathrm{s}$ is a fixed design constant and
not an optimization variable. Accordingly, the discrete-domain design vector is
$\Theta_\mathrm{d}=\{\alpha,\beta\}$ throughout Sections~\ref{sec3}--\ref{sec4}; $T_\mathrm{s}$ is
listed in $\Theta_\mathrm{d}$ only in the statement of Theorem~\ref{thm:coupling_non_unique}.

\subsection{System Open-loop Bode Plot}
The system open-loop transfer function is depicted as follows:
\begin{equation}
    G_{\mathrm{ol}}(z) = \mathcal{Z}\!\left\{ G_{\mathrm{pwm}}(s) \cdot G_{\mathrm{filt}}(s) \right\} \cdot G_{\mathrm{c}}(z) \cdot G_{\mathrm{delay}}(z), \label{eq:sys_ol}
\end{equation}
where $\mathcal{Z}\{\cdot\}$ denotes the $\mathcal{Z}$-transform.

\begin{figure}[htbp]
  \centering
  % line 1
  \begin{minipage}[t]{0.48\linewidth}
    \centering
    \includegraphics[width=\linewidth]{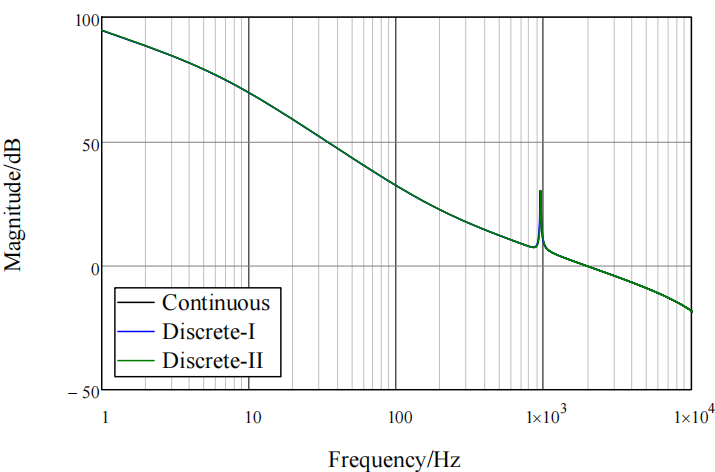}
    \centerline{\footnotesize (a) Magnitude.}
  \end{minipage}
  \hfill
  \begin{minipage}[t]{0.48\linewidth}
    \centering
    \includegraphics[width=\linewidth]{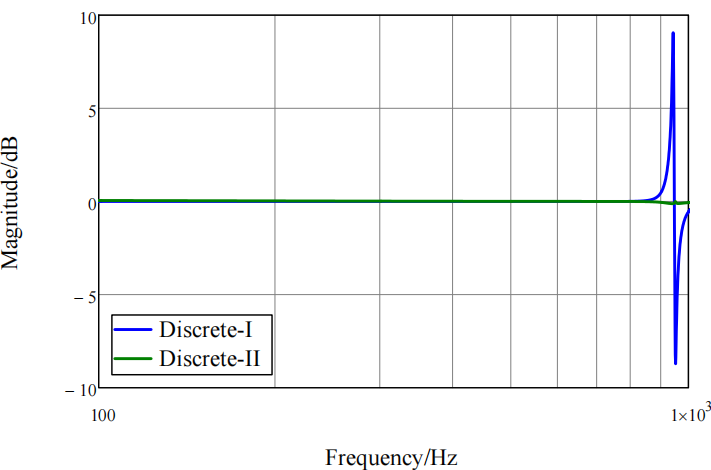}
    \centerline{\footnotesize (b) Magnitude error.}
  \end{minipage}

  \vspace{0.5em}

  % line 2
  \begin{minipage}[t]{0.48\linewidth}
    \centering
    \includegraphics[width=\linewidth]{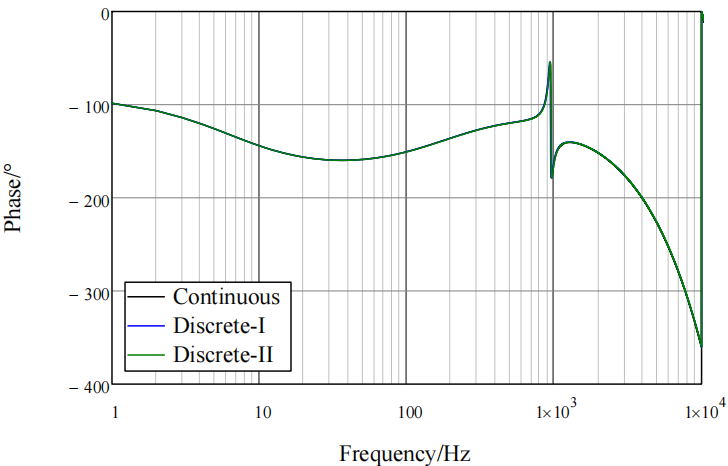}
    \centerline{\footnotesize (c) Phase.}
  \end{minipage}
  \hfill
  \begin{minipage}[t]{0.48\linewidth}
    \centering
    \includegraphics[width=\linewidth]{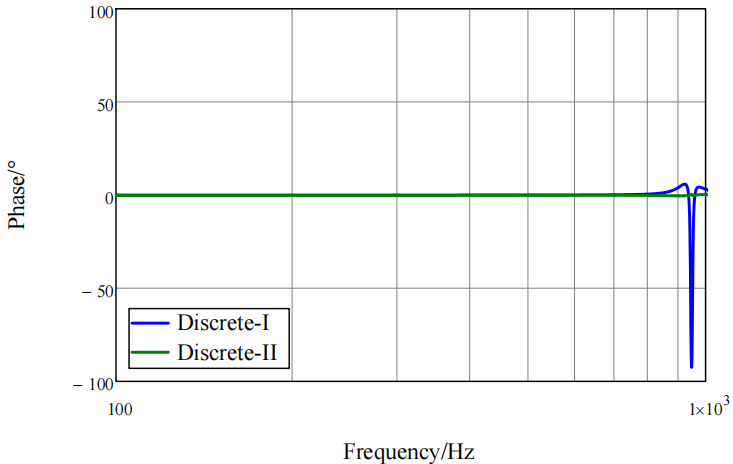}
    \centerline{\footnotesize (d) Phase error.}
  \end{minipage}

  \caption{System open-loop Bode plot.}
  \label{fig:sys_bode_plot}
\end{figure}

Fig.~\ref{fig:sys_bode_plot} illustrates the system Bode plots for a resonant frequency of 950 Hz, 
obtained using the SBT method with a fixed shape factor of $\alpha = 0.5$.
Fig.~\ref{fig:sys_bode_plot}(a) compares the magnitude responses: 
the continuous-domain reference (black solid line), 
the imprecisely matched discrete system (Discrete-I, blue solid line, $\beta=1$), 
and the precisely matched discrete system (Discrete-II, green solid line, 
$\beta=1.007$ prewarping factor). 
Although the three curves nearly overlap in the main plot, 
the magnified view in Fig.~\ref{fig:sys_bode_plot}(b) reveals the 
magnitude deviation from the continuous-domain reference. 
Specifically, Discrete-I exhibits a significant magnitude error of 
approximately 9~dB around the resonant frequency, 
whereas Discrete-II maintains near-zero error.
Fig.~\ref{fig:sys_bode_plot}(c) and Fig.~\ref{fig:sys_bode_plot}(d) present the phase response and the corresponding phase error, respectively. 
Similarly, Discrete-I suffers from severe phase lag (up to $-90^\circ$) at resonance, 
while Discrete-II accurately tracks the continuous-domain phase characteristic.
These results underscore that improper selection of discretization parameters ($\beta$) 
induces substantial frequency warping in the resonant stage, 
necessitating careful co-design of discretization parameters 
for high-precision digital implementation.

\subsection{Formal Definition of the Joint Optimization Problem}
\newtheorem{theorem}{Theorem}
\newtheorem{proof}{Proof}
\newtheorem{remark}{Remark} % added for the remarks supporting the revised theorem

\begin{theorem}[Non-Uniqueness and Directional Coupling under Continuous-Discrete Parameterization]
\label{thm:coupling_non_unique}
For the PIR controller discretized via SBT, let
$\mathcal{F}(\Theta_\mathrm{c},\Theta_\mathrm{d})$ denote the resulting discrete-time open-loop frequency
response, with
$\Theta_\mathrm{c}=\{K_\mathrm{p},\tau_\mathrm{i},K_\mathrm{r},\omega_\mathrm{b},\omega_n\}$ and
$\Theta_\mathrm{d}=\{\alpha,\beta,T_\mathrm{s}\}$. Then:
\begin{enumerate}
\item[(a)] \emph{(Non-uniqueness.)} $\mathcal{F}$ is not injective: there exists a continuous
one-parameter family of distinct tuples $(\Theta_\mathrm{c},\Theta_\mathrm{d})\neq(\Theta_\mathrm{c}',\Theta_\mathrm{d}')$
with $\mathcal{F}(\Theta_\mathrm{c},\Theta_\mathrm{d})=\mathcal{F}(\Theta_\mathrm{c}',\Theta_\mathrm{d}')$.
\item[(b)] \emph{(Directional sharpness.)} The redundancy is intrinsically dual-domain: freezing
either domain alone removes it. Specifically, for a given response,
\begin{itemize}
\item[(i)] if $\Theta_\mathrm{d}$ is fixed, the continuous parameters are generically unique;
\item[(ii)] if $\Theta_\mathrm{c}$ is fixed, $(\alpha,\beta)$ are generically unique.
\end{itemize}
\end{enumerate}
\end{theorem}

\begin{proof}
\textbf{Step 1 (dimension counting).}
The eight parameters $(K_\mathrm{p},\tau_\mathrm{i},K_\mathrm{r},\omega_\mathrm{b},\omega_n,\alpha,\beta,T_\mathrm{s})$
enter the model only through the seven discrete coefficients
$(A_0,A_1,C_0,C_1,C_2,D_1,D_2)$ of \eqref{eq:disc_pir}--\eqref{eq:qr_coeff_2}, 
and the plant in \eqref{eq:sys_ol} is independent of them. 
Hence
$\mathcal{F}=\mathcal{G}\circ\mathcal{C}$ factorizes through the coefficient map
$\mathcal{C}:\mathbb{R}^{8}\to\mathbb{R}^{7}$, so $\ker$-type redundancy is unavoidable:
fibers of $\mathcal{C}$ are generically at least one-dimensional.
Because both the objective \eqref{eq:unified_obj} and every stability constraint of
Section~\ref{sec:opt_constraints} depend on $\bm{\theta}$ only through these seven coefficients,
the optimization landscape necessarily contains flat directions along which $J$ is constant.

\textbf{Step 2 (explicit equivalence family).}
Consider, for any $k>0$, $k\neq 1$, the joint scaling
\begin{equation}
  \begin{split}
  &\omega_n'=k\omega_n,\quad \omega_\mathrm{b}'=k\omega_\mathrm{b},\quad \tau_\mathrm{i}'=\tau_\mathrm{i}/k,\\
  &\beta'=\beta/k, \quad K_\mathrm{p}'=K_\mathrm{p},\quad K_\mathrm{r}'=K_\mathrm{r}.
  \end{split}
\label{eq:scaling_group}
\end{equation}
Substituting into $\Psi_1$ gives
\[
\Psi_1'=\frac{k^2}{(\beta T_\mathrm{s})^2}+\frac{2(k\omega_\mathrm{b})\alpha}{(\beta/k)T_\mathrm{s}}+k^2\omega_n^2\alpha^2=k^2\Psi_1,
\]
and, by the same homogeneity, $\Psi_2'=k^2\Psi_2$ and $\Psi_3'=k^2\Psi_3$. Therefore
$D_1'=\Psi_2'/\Psi_1'=D_1$ and $D_2'=D_2$; moreover
\[
C_0'=\frac{2K_\mathrm{r}(k\omega_\mathrm{b})\alpha}{(\beta/k)T_\mathrm{s}\,k^2\Psi_1}
=\frac{2K_\mathrm{r}\omega_\mathrm{b}\alpha}{\beta T_\mathrm{s}\Psi_1}=C_0,
\]
and identically $C_1'=C_1$, $C_2'=C_2$. Finally, since
$\beta'T_\mathrm{s}/\tau_\mathrm{i}'=(\beta/k)T_\mathrm{s}\,k/\tau_\mathrm{i}=\beta T_\mathrm{s}/\tau_\mathrm{i}$,
the PI branch is invariant as well: $A_0'=A_0$, $A_1'=A_1$.
Thus all seven coefficients, and hence $\mathcal{F}$, are invariant under \eqref{eq:scaling_group},
which proves (a).
Two of the four substitutions in \eqref{eq:scaling_group} are essential and were missing from a naive
construction: without $\omega_\mathrm{b}'=k\omega_\mathrm{b}$ the term $2\omega_\mathrm{b}\alpha/(\beta T_\mathrm{s})$
scales by $k$ rather than $k^2$, and without $\tau_\mathrm{i}'=\tau_\mathrm{i}/k$ the coefficients
$A_0,A_1$ are not preserved.

\textbf{Step 3(i) ($\Theta_\mathrm{d}$ fixed $\Rightarrow$ $\Theta_\mathrm{c}$ unique).}
With $(\alpha,\beta,T_\mathrm{s})$ given, \eqref{eq:disc_pir} gives
$A_0+A_1=K_\mathrm{p}\beta T_\mathrm{s}/\tau_\mathrm{i}$ and
$A_0=K_\mathrm{p}+\alpha(A_0+A_1)$, hence
$K_\mathrm{p}=A_0-\alpha(A_0+A_1)$ and
$\tau_\mathrm{i}=\beta T_\mathrm{s}K_\mathrm{p}/(A_0+A_1)$ are determined.
Thereafter $D_1,D_2$ are two equations in the two unknowns $(\omega_\mathrm{b},\omega_n)$ and
$C_0$ fixes $K_\mathrm{r}$; the solution is unique except on a set of measure zero
(e.g., degenerate $\omega_\mathrm{b}\to0$).

\textbf{Step 3(ii) ($\Theta_\mathrm{c}$ fixed $\Rightarrow$ $(\alpha,\beta)$ unique).}
With $\Theta_\mathrm{c}$ given, \eqref{eq:disc_pir} yields the two constants
$\alpha\beta=(A_0/K_\mathrm{p}-1)\tau_\mathrm{i}/T_\mathrm{s}$ and
$(1-\alpha)\beta=(A_1/K_\mathrm{p}+1)\tau_\mathrm{i}/T_\mathrm{s}$.
Adding them determines $\beta$ uniquely, and then $\alpha$ is obtained from their ratio.
Hence no two distinct $(\alpha,\beta)$ pairs can reproduce the same controller once
$\Theta_\mathrm{c}$ is frozen, which proves (b).
\end{proof}

Theorem \ref{thm:coupling_non_unique} reveals that the discrete frequency response is
jointly determined by $\Theta_\mathrm{c}$ and $\Theta_\mathrm{d}$, not independently by either.
Serial design workflows freeze one prioritized domain and 
therefore restrict the search to a slice
that does not contain the joint optimum.
We therefore distinguish between two approaches: 
\textbf{Separate Optimization}, which prioritizes implementability over performance, 
and \textbf{Joint Optimization}, which targets the best attainable performance. 
The subsequent sections elaborate on these methodologies via the analytical EFDAM (Section~\ref{sec3}) and 
the numerical ICASM (Section~\ref{sec4}).

%-------------------------------------------------------------------------------
% Section III: Explicit Frequency-Domain Analytical Method (EFDAM)
%-------------------------------------------------------------------------------
\section{Explicit Frequency-Domain Analytical Method (EFDAM)}\label{sec3}
\subsection{Theory of EFDAM} 
\begin{figure}[htbp]
  \centering
  % line 1
  \begin{minipage}[t]{0.50\linewidth}
    \centering
    \includegraphics[width=\linewidth]{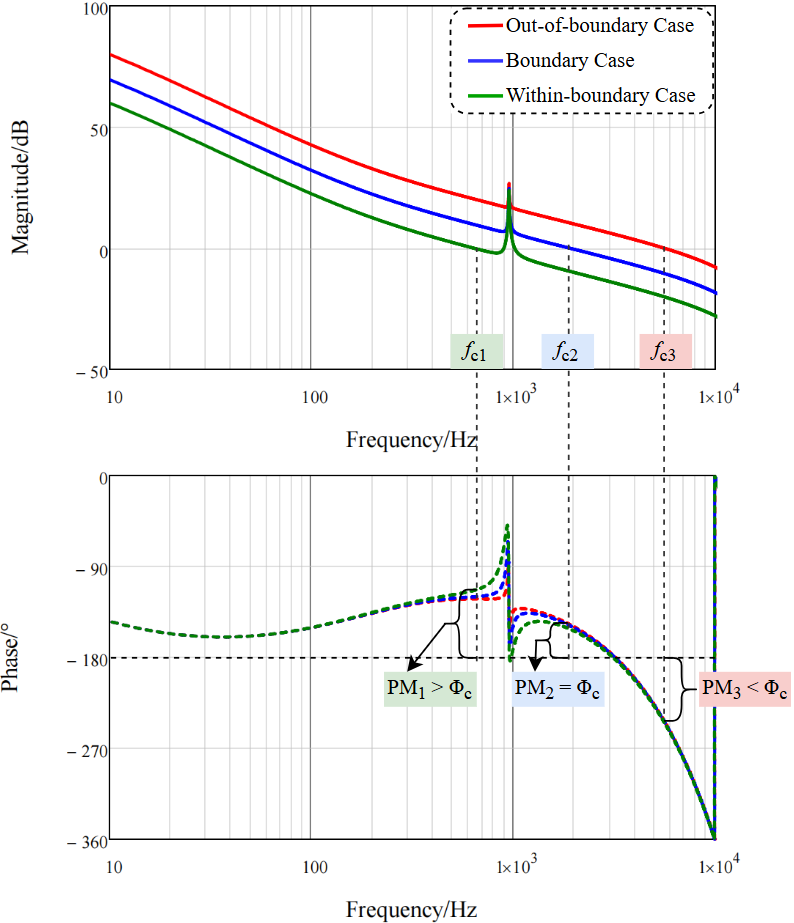}
    \centerline{\footnotesize (a) Phase margin vs. $K_{\mathrm{p}}$.}
  \end{minipage}
  \hfill
  \begin{minipage}[t]{0.46\linewidth}
    \centering
    \includegraphics[width=\linewidth]{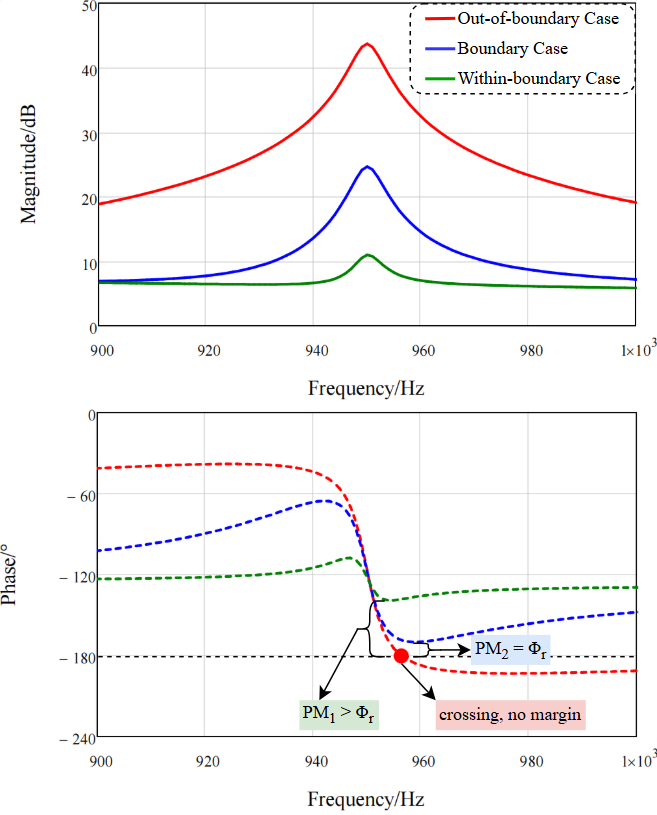}
    \centerline{\footnotesize (b) Phase margin vs. $K_{\mathrm{r}}$.}
  \end{minipage}
  \caption{Open-loop Bode plots showing the phase-margin boundary cases versus (a) $K_{\mathrm{p}}$ and (b) $K_{\mathrm{r}}$.}
  \label{fig:efdam}
\end{figure}

The design principles for optimized parameter tuning are established upon the rigorous \emph{logarithmic frequency stability criterion} 
for continuous-domain parameters, 
and the minimization of discretization errors for discrete-domain parameters based on Taylor-series error bound analysis and 
discrete frequency response alignment—derivable entirely in closed form without any recourse to numerical search tools.
Specifically, the design rationale is based on the following criteria:

\begin{enumerate}
    \item The optimized values for $K_{\mathrm{p}}$ and $\tau_{\mathrm{i}}$ are achieved when $K_{\mathrm{p}}$ is maximized and 
    $\tau_{\mathrm{i}}$ is minimized, respectively. 
    As established in \cite{holmes_tpel_2009}, 
    this extremum occurs concurrently with the maximization of the crossover frequency. 
    Under this condition, the phase margin is precisely equal to $\Phi_\mathrm{c}$, 
    as defined in Fig.~\ref{fig:efdam}(a) (Boundary case).
    
    \item The optimized value for $K_{\mathrm{r}}$ corresponds to its maximum achievable gain. 
    Based on the logarithmic frequency stability criterion, 
    this maximum $K_{\mathrm{r}}$ is attained when the phase curve avoids crossing $-180^\circ$ 
    while maintaining a specific resonant phase margin, denoted as $\Phi_\mathrm{r}$, 
    as illustrated in Fig.~\ref{fig:efdam}(b) (Boundary case).

    \item The optimized value for $\omega_{\mathrm{b}}$ is its minimum. 
    This minimum value is determined by the requirement to maintain a specified magnitude ratio (e.g., 60\% of the peak resonant magnitude) 
    across the anticipated grid frequency fluctuation range, 
    ensuring robust harmonic rejection under grid frequency deviations.

    \item The optimized $\alpha$ is selected to minimize the discretization error, as quantified by the error-bound analysis of the SBT.

    \item The optimized $\beta$ is the one that eliminates frequency warping at $\omega_n$ based on the SBT-discretized 
    frequency characteristic.
\end{enumerate}

\subsection{Derivation of Explicit Formulas for Optimized Continuous-Domain Parameters}
\subsubsection{Derivation of optimized $K_{\mathrm{p}}$ and $\tau_{\mathrm{i}}$ ($K_{\mathrm{p\_opt}}$ and $\tau_{\mathrm{i\_opt}}$)} 
As illustrated in Fig.~\ref{fig:efdam}(a), 
the proportional gain $K_{\mathrm{p}}$ dictates the crossover frequency $f_{\mathrm{c}}$.
The figure contrasts three cases against a target phase margin $\Phi_{\mathrm{c}}$:
\begin{itemize}
    \item Within-boundary case ($K_{\mathrm{p}}$ is insufficient, green curve): 
    Represents a conservative design with $f_{\mathrm{c1}}$ and $\mathrm{PM}_1 > \Phi_{\mathrm{c}}$.
    While stable, it sacrifices dynamic performance.
    
    \item Boundary case ($K_{\mathrm{p}}$ is optimal, blue curve): 
    Represents the critical boundary where $f_{\mathrm{c2}}$ yields $\mathrm{PM}_2 = \Phi_{\mathrm{c}}$. 
    This corresponds to the maximum admissible $K_{\mathrm{p}}$ and defines the optimal design point.

    \item Out-of-boundary case ($K_{\mathrm{p}}$ is excessive, red curve): 
    Results in a crossover frequency $f_{\mathrm{c3}}$ 
    where the phase margin $\mathrm{PM}_3 < \Phi_{\mathrm{c}}$, violating the stability requirement.

    \end{itemize}

Consequently, as established by the logarithmic frequency stability criterion \cite{holmes_tpel_2009}, 
the maximum proportional gain $K_{\mathrm{p\_max}}$ and minimum integrator time constant $\tau_{\mathrm{i\_min}}$ are derived 
to operate at this boundary. 
The derivation assumes: (1) the load resistance $R_{\mathrm{L}}$ is neglected, and (2) $\omega_{\mathrm{c}} \tau_{\mathrm{i}} \gg 1$.

Under these conditions, the maximum attainable crossover angular frequency is approximated as:
\begin{equation}
\omega_{\mathrm{c\_max}} \approx \frac{\pi/2 - \Phi_{\mathrm{c}} - \Delta_\Phi(n)}{1.5 \, T_{\mathrm{s}}}.
\label{eq:omega_c_max}
\end{equation}
where $n$ is the harmonic order addressed by the resonant branch
($n=19$ in this work; $n=0$ denotes a PI regulator with no resonant branch), 
and $\Delta_\Phi(n)$ denotes the additional phase lag at $\omega_\mathrm{c}$. 

This lag is the sum of the integral-action lag and the phase
contribution of the resonant tail at $\omega_\mathrm{c}$, 
and it admits the closed-form decomposition
\begin{equation}
\Delta_\Phi \approx \underbrace{\arctan\frac{1}{\omega_\mathrm{c}\tau_\mathrm{i}}}_{\text{integral action}}
\;+\;
\underbrace{\arctan\frac{2K_\mathrm{r}\omega_\mathrm{b}\omega_\mathrm{c}}
{K_\mathrm{p}\left(\omega_\mathrm{c}^{2}-\omega_n^{2}\right)}}_{\text{resonant tail }(\omega_\mathrm{c}>\omega_n)},
\label{eq:dphi_decomp}
\end{equation}
which follows directly from \eqref{eq:sys_ol_ph_app} by evaluating the imaginary/real ratio of the
controller at $\omega_\mathrm{c}$. At the design point of Table~\ref{tab:test_system_params} it yields
$5.71^\circ+1.79^\circ\approx7.5^\circ$.
Since $K_\mathrm{r}$ and $\omega_\mathrm{b}$ appear in \eqref{eq:dphi_decomp}, substituting it into
\eqref{eq:omega_c_max} would make the crossover-frequency estimate implicit. To keep the method strictly
closed-form, $\Delta_\Phi(n)$ is therefore replaced by the fitted bound
$\Delta_\Phi(n) \approx \mathrm{round}\bigl[ 2 \cdot \min(5.5 + 0.125(n-3), 7.0) \bigr] \frac{\pi}{360},$
i.e., by at most $7^\circ$ for $n\ge15$, which for the present design point underestimates
\eqref{eq:dphi_decomp} by $\approx0.5^\circ$ and thus overestimates $\omega_{\mathrm{c\_max}}$ by less
than $1\%$. 
Whenever a certain phase margin is required, \eqref{eq:dphi_decomp} can be used inside a
single fixed-point iteration on \eqref{eq:omega_c_max} instead of the fitted bound; this residual
optimism is one of the reasons why EFDAM is used as a physics-informed initialization whose feasibility
is finally certified by the exact numerical constraints of ICASM.

The corresponding maximum proportional gain is given by:
\begin{equation}
K_{\mathrm{p\_max}} \approx \omega_{\mathrm{c\_max}} L_{\mathrm{f}}. \label{eq:Kp_max}
\end{equation}

Furthermore, the minimum integrator time constant, 
ensuring the integral action does not interfere with the phase margin, is approximated as:
\begin{equation}
\tau_{\mathrm{i\_min}} \approx \frac{10}{\omega_{\mathrm{c\_max}}}. \label{eq:tau_i_min}
\end{equation}

The factor $10$ is a phase-budget choice rather than a heuristic: it fixes
$\kappa\triangleq\omega_\mathrm{c}\tau_\mathrm{i}=10$, so the integral action contributes exactly
$\arctan(1/\kappa)=5.71^\circ$ of lag at crossover. Any other $\kappa$ may be selected with the same
transparency ($\kappa=20\Rightarrow2.86^\circ$, $\kappa=5\Rightarrow11.3^\circ$), the trade-off being
between the crossover-frequency budget and the speed of the low-frequency dynamics.

Consequently, the optimized parameters are assigned as:
\begin{equation}
K_{\mathrm{p\_opt}} = K_{\mathrm{p\_max}}, \quad \tau_{\mathrm{i\_opt}} = \tau_{\mathrm{i\_min}}. \label{eq:Kp_tau_i_opt}
\end{equation}

\subsubsection{Derivation of optimized $K_{\mathrm{r}}$ and $\omega_{\mathrm{b}}$ ($K_{\mathrm{r\_opt}}$ and $\omega_{\mathrm{b\_opt}}$)} 
Fig.~\ref{fig:efdam}(b) illustrates the determination of the maximum admissible resonant gain $K_{\mathrm{r}}$ 
by comparing three cases against a target resonant phase margin $\Phi_{\mathrm{r}}$:
\begin{itemize}
    \item Within-boundary case ($K_{\mathrm{r}}$ is insufficient, green curve):
    Represents a conservative design where the phase margin $\mathrm{PM}_1 > \Phi_{\mathrm{r}}$. 
    While stable, it sacrifices disturbance rejection performance at resonance.
    \item Boundary case ($K_{\mathrm{r}}$ is optimal, blue curve): 
    Represents the critical boundary where $\mathrm{PM}_2 = \Phi_{\mathrm{r}}$. 
    This corresponds to the maximum admissible $K_{\mathrm{r}}$ and defines the optimal design point.
    \item Out-of-boundary case ($K_{\mathrm{r}}$ is excessive, red curve): 
    Causes the phase to cross $-180^\circ$, violating the stability requirement.
\end{itemize}

Decomposing the open-loop transfer function into real and imaginary parts yields 
the linear magnitude $M(\omega)$, its decibel representation
$L(\omega)\triangleq20\log_{10}M(\omega)$ (used consistently from here on), 
and the phase $\varphi(\omega)$.

\begin{align}
    M(\omega) &= \sqrt{\left[\operatorname{Re}(G_{\mathrm{ol}})\right]^2 + \left[\operatorname{Im}(G_{\mathrm{ol}})\right]^2} \\
    \varphi(\omega) &= \arctan\left( \frac{\operatorname{Im}(G_{\mathrm{ol}})}{\operatorname{Re}(G_{\mathrm{ol}})} \right) \\
    L(\omega) &= 20\log_{10} M(\omega)
\end{align}

Specifically, for the PIR controller analyzed, $L(\omega)$ and $\varphi(\omega)$ take the form in
Eqs.~\eqref{eq:sys_ol_mag_app} and \eqref{eq:sys_ol_ph_app}, obtained with $R_{\mathrm{L}}$ neglected and the ZOH modeled as
$e^{-0.5sT_{\mathrm{s}}}$ \cite{holmes_tpel_2009}.

\begin{equation}
L(\omega) = 20\log_{10}{\biggl(\tfrac{\sqrt{\left( K_{\mathrm{p}} + \tfrac{2 K_{\mathrm{r}} \omega_{\mathrm{b}} \omega B}{A^2 + B^2} \right)^2 + \left( \tfrac{-K_{\mathrm{p}}}{\tau_{\mathrm{i}} \omega} + \tfrac{2 K_{\mathrm{r}} \omega_{\mathrm{b}} \omega A}{A^2 + B^2} \right)^2}}{\omega L_{\mathrm{f}}} \biggr)},
\label{eq:sys_ol_mag_app}
\end{equation}

\begin{equation}
\varphi(\omega) = \arctan\biggl(\frac{\frac{-K_{\mathrm{p}}}{\tau_{\mathrm{i}} \omega} + \frac{2 K_{\mathrm{r}} \omega_{\mathrm{b}} \omega A}{A^2 + B^2}}{K_{\mathrm{p}} + \frac{2 K_{\mathrm{r}} \omega_{\mathrm{b}} \omega B}{A^2 + B^2}}\biggr) - 1.5\omega T_{\mathrm{s}} - \frac{\pi}{2}.
\label{eq:sys_ol_ph_app}
\end{equation}
where $A = \omega_n^2 - \omega^2$, $B = 2 \omega_{\mathrm{b}} \omega$.

To accurately determine the phase dip around resonance, 
the phase characteristic of the controller $\varphi_{\mathrm{c}}(\omega)$ is formulated as:
\begin{equation}
    \varphi_{\mathrm{c}}(\omega) = \arctan\biggl( \frac{N_\phi(\omega)}{D_\phi(\omega)} \biggr)
\end{equation}
where $N_\phi(\omega) = -K_{\mathrm{p}} (A^2 + B^2) + 2K_{\mathrm{r}} \omega_{\mathrm{b}} \tau_{\mathrm{i}} \omega^2 A$ and
 $D_\phi(\omega) = \omega \tau_{\mathrm{i}} [K_{\mathrm{p}} (A^2+B^2) + 2K_{\mathrm{r}} \omega_{\mathrm{b}} \omega B]$.

The phase peak (local extremum) occurs when the derivative of the total phase equals zero. 
Considering that the delay term is approximately $-1.5\omega_n T_{\mathrm{s}}$ (since the phase peak occurs in the vicinity of $\omega_n$), 
the core condition simplifies to:
\begin{equation}
    N_\phi'(\omega) D_\phi(\omega) - N_\phi(\omega) D_\phi'(\omega) = 0. \label{eq:omega_r_max}
\end{equation}

Analysis shows that the phase peak frequency $\omega_{\mathrm{r\_max}}$ deviates from $\omega_n$. 
Solving this transcendental equation yields the precise $\omega_{\mathrm{r\_max}}$ required for calculating $\mathrm{PM}$. 
However, a closed-form analytical solution is intractable; numerical solvers are recommended. 
Substituting $\omega_{\mathrm{r\_max}}$ back into the phase function \eqref{eq:sys_ol_ph_app} yields the phase margin constraint:
\begin{equation}
    \arctan\left( \frac{N_\phi(\omega_{\mathrm{r\_max}})}{D_\phi(\omega_{\mathrm{r\_max}})} \right)  \approx - \frac{\pi}{2} + 1.5\omega_n T_{\mathrm{s}} + \Phi_{\mathrm{r}}.\label{eq:resonant_boundary_cond_1}
\end{equation}

The minimum value of $\omega_{\mathrm{b}}$ is determined by the requirement to maintain 
a specified magnitude ratio $\eta$ (e.g., $\eta=0.6$ means that the loop magnitude at the
edges of the band must be at least $60\%$ of the peak magnitude)
across the anticipated resonant frequency fluctuation range, denoted as $[\omega_{n-}, \omega_{n+}]$,
\begin{equation}
    \min\{L(\omega_{n-}),\,L(\omega_{n+})\} = \eta L(\omega_n).
    \label{eq:resonant_boundary_cond_2}
\end{equation}

The boundary pair $(K_{\mathrm{r\_max}}, \omega_{\mathrm{b\_min}})$ is obtained by 
solving the phase margin constraint (Eq.~\eqref{eq:resonant_boundary_cond_1}) 
and the magnitude constraint (Eq.~\eqref{eq:resonant_boundary_cond_2}) simultaneously, 
where $K_{\mathrm{p}}$ and $\tau_{\mathrm{i}}$ are fixed at their optimal values 
defined in Eq.~\eqref{eq:Kp_tau_i_opt}. 
This solution corresponds to the boundary case ensuring $\mathrm{PM} = \Phi_{\mathrm{r}}$.

Consequently, the optimized parameters are assigned as:
\begin{equation}
K_{\mathrm{r\_opt}} = K_{\mathrm{r\_max}}, \quad \omega_{\mathrm{b\_opt}} = \omega_{\mathrm{b\_min}}. \label{eq:Kr_omega_b_opt}
\end{equation}

\subsection{Derivation of Explicit Formulas for Optimized Discrete-Domain Parameters}
\subsubsection{Derivation of optimized $\alpha$ ($\alpha_{\mathrm{opt}}$)} 
To rigorously establish the relationship between $\alpha$ and the inherent regulator-induced damping, 
we analyze the Taylor series expansion of the SBT mapping. 
Let $z_{\mathrm{exact}} = e^{\beta s T_{\mathrm{s}}}$ 
be the exact exponential mapping and $z_{\mathrm{app}} = (1+(1-\alpha)\beta s T_{\mathrm{s}})/(1-\alpha \beta s T_{\mathrm{s}})$ 
be the SBT approximation. Their expansions are:

\begin{equation}
z_{\mathrm{exact}} = 1 + \beta s T_{\mathrm{s}} + \frac{(\beta s T_{\mathrm{s}})^2}{2} + \frac{(\beta s T_{\mathrm{s}})^3}{6} + O((\beta s T_{\mathrm{s}})^4),
\end{equation}
\begin{equation}
\begin{aligned}
z_{\mathrm{app}} &= \frac{1+(1-\alpha)\beta s T_{\mathrm{s}}}{1-\alpha \beta s T_{\mathrm{s}}} = \left[1+(1-\alpha)\beta s T_{\mathrm{s}}\right] \sum_{k=0}^{\infty} (\alpha \beta s T_{\mathrm{s}})^k \\
                 &= 1 + \beta s T_{\mathrm{s}} + \alpha (\beta s T_{\mathrm{s}})^2 + \alpha^2 (\beta s T_{\mathrm{s}})^3 + O((\beta s T_{\mathrm{s}})^4).
\end{aligned}
\end{equation}

The error $\epsilon  = z_{\mathrm{exact}} - z_{\mathrm{app}}$ is:
\begin{equation}
\epsilon = \underbrace{\left(\tfrac{1}{2} - \alpha\right)}_{\text{Quadratic coeff.}} (\beta s T_{\mathrm{s}})^2 + \underbrace{\left(\tfrac{1}{6} - \alpha^2\right)}_{\text{Cubic coeff.}} (\beta s T_{\mathrm{s}})^3 + O((\beta s T_{\mathrm{s}})^4).
\end{equation}

The quadratic error term is eliminated when $\alpha = 0.5$,
thus,
\begin{equation}
\alpha_{\mathrm{opt}} = 0.5. \label{eq:alpha_opt}
\end{equation}

\subsubsection{Derivation of optimized $\beta$ ($\beta_{\mathrm{opt}}$)} 
The optimized $\beta$ is derived by enforcing a zero-warping condition at the resonant frequency $\omega_n$. 
Substituting the mapping $s=j\omega_n \mapsto z=e^{j\omega_n T_{\mathrm{s}}}$ into the SBT transformation \eqref{eq:sbt} 
with $\alpha=0.5$ yields:

\begin{equation}
j\omega_n = \frac{1}{\beta T_{\mathrm{s}}} \frac{e^{j\omega_n T_{\mathrm{s}}}-1}{0.5(e^{j\omega_n T_{\mathrm{s}}}+1)}.
\label{eq:beta_derivation_start}
\end{equation}

Applying the identities $e^{j\theta}-1 = j e^{j\theta/2} \cdot 2\sin(\theta/2)$ and $e^{j\theta}+1 = e^{j\theta/2} \cdot 2\cos(\theta/2)$, 
the fraction simplifies to $j\tan(\theta/2)$. Thus,

\begin{equation}
j\omega_n = \frac{2}{\beta T_{\mathrm{s}}} j \tan\left(\frac{\omega_n T_{\mathrm{s}}}{2}\right).
\end{equation}

Canceling $j$ and solving for $\beta$ gives the $\beta_\mathrm{opt}$:

\begin{equation}
\beta_{\mathrm{opt}} = \frac{2 \tan\left(\frac{\omega_n T_{\mathrm{s}}}{2}\right)}{\omega_n T_{\mathrm{s}}}
\label{eq:beta_opt}
\end{equation}

The SBT-discretized PIR controller admits closed-form solutions 
given by Eqs.~\eqref{eq:Kp_tau_i_opt}, \eqref{eq:Kr_omega_b_opt}, 
\eqref{eq:alpha_opt}, and \eqref{eq:beta_opt}, 
which makes the loop-shaping and discretization steps mutually consistent without any
numerical search; note that $\beta_\mathrm{opt}$ depends on $\omega_n$, i.e., on a continuous-domain
parameter, so the two steps are linked rather than independent.
Optimization based on Table~\ref{tab:test_system_params} yields:
$K_{\mathrm{p}}=3.022$, $\tau_{\mathrm{i}}=0.811\,\mathrm{ms}$ ($\Phi_{\mathrm{c}}=30^\circ$);
$K_{\mathrm{r}}=22.36$, $\omega_{\mathrm{b}}=19.96\,\mathrm{rad/s}$ ($\Phi_{\mathrm{r}}=10^\circ$);
$\alpha=0.5$, $\beta=1.0075$.

\begin{table}[htbp]
\centering
\caption{Circuit Parameters of The Test System}
\label{tab:test_system_params}
\begin{tabular}{|l|l|}
\hline
\textbf{Circuit Parameter} & \textbf{Value} \\
\hline
Filter inductance $L_{\mathrm{f}}$  & 245~$\mu$H \\
Series resistance $R_{\mathrm{L}}$  & $0.03~\Omega$ \\
Sampling period $T_{\mathrm{s}}$    & 50~$\mu$s \\
DC bus voltage ($U_{\mathrm{dc}}$)  & $500~\mathrm{V}$ \\
Grid voltage                                          & 220~$V_{\mathrm{rms}}$ \\
Grid frequency                                        & 50~Hz \\
Grid frequency range                                  & 49.5--50.5~Hz   \\
Test resonant frequency $f_n$                         & 950~Hz          \\
Resonant frequency range $[f_{n-}, f_{n+}]$           & 940.5--959.5~Hz \\
Resonant magnitude maintenance ratio $\eta$           & 0.6             \\
\hline
\end{tabular}
\end{table}

%-------------------------------------------------------------------------------
% Section IV: Implicit Computer-Aided Search Method (ICASM)
%-------------------------------------------------------------------------------
\section{Implicit Computer-Aided Search Method (ICASM)}\label{sec4}
\subsection{Construction of Unified Objective Function}
Controller tuning is formulated as a multi-objective optimization over the open-loop model $G_{\mathrm{ol}}(z)$, 
obtained by substituting $s=j\omega$ and $z=e^{j\omega T_\mathrm{s}}$ into \eqref{eq:sys_ol}. 
Three frequency-domain metrics are prioritized:
\begin{enumerate}
\item \textbf{Bandwidth}: crossover frequency $\omega_\mathrm{c}$, defined by $|G_{\mathrm{ol}}(j\omega_\mathrm{c})|=1$;
\item \textbf{Resonant Magnitude}: $L_n = 20\log_{10}|G_{\mathrm{ol}}(j\omega_n)|$;
\item \textbf{Fundamental Magnitude}: $L_1 = 20\log_{10}|G_{\mathrm{ol}}(j\omega_1)|$.
\end{enumerate}
Each metric is normalized by its respective upper bound ($\omega_{\mathrm{c\_max}}$, $L_{n\_\max}$, $L_{1\_\max}$) 
derived from feasibility analysis or prior scans. 
Then the objective is given by:
\begin{equation}
J(\bm{\theta})=
\lambda_1\frac{L_{1}}{L_{1\_\max}}+
\lambda_2\frac{L_{n}}{L_{n\_\max}}+
\lambda_3\frac{\omega_{\mathrm{c}}}{\omega_{\mathrm{c\_max}}},
\label{eq:unified_obj}
\end{equation}
where $\bm{\theta}=[K_\mathrm{p}, \tau_\mathrm{i}, \omega_\mathrm{b}, K_\mathrm{r}, \alpha, \beta]^{\mathsf{T}}$ 
($\omega_n$ and $T_{\mathrm{s}}$ are pre-defined in design specifications). 
The bounds $\omega_{\mathrm{c\_max}}$, $L_{n\_\max}$, and $L_{1\_\max}$ represent 
the maximum achievable value for each metric when optimized individually.
The tuning problem is thus formally stated as:
\begin{align}
&\max_{\bm{\theta}} \; J(\bm{\theta}) \notag \\
&\text{s.t.} \quad \text{Constraints in Section~\ref{sec:opt_constraints},}
\end{align}
which guarantees robust stability while enforcing the desired performance trade-off.

\paragraph{Problem dimensionality}
For the PIR regulator considered here the search vector is
$\bm{\theta}\in\mathbb{R}^{6}$ ($K_\mathrm{p},\tau_\mathrm{i},K_\mathrm{r},\omega_\mathrm{b},\alpha,\beta$),
with $\omega_n$ and $T_\mathrm{s}$ fixed by the specification and the hardware. The difficulty therefore
does not lie in the number of variables; it lies in the non-convexity of the search space
(the resonant peak makes the phase a non-monotone function of $\omega$ and of
$\omega_\mathrm{b}$) and non-smoothness ($\omega_\mathrm{c}$ and $\omega_\phi$ are
$\arg$-type quantities, so $J$ is only piecewise smooth).

\paragraph{Choice and role of the weights}
Each metric in \eqref{eq:unified_obj} is normalized by the maximum value it can attain when optimized
alone, so every term is dimensionless and of order unity; the weights therefore express a pure priority
ordering and are not needed to fix units or scales. The neutral choice $\lambda_1=\lambda_2=\lambda_3=1/3$
is adopted as the baseline, and the framework allows any other preference to be imposed without
re-deriving the model.

\subsection{Analysis of Constraint Conditions}\label{sec:opt_constraints}
Based on $G_{\mathrm{ol}}(z)$ in \eqref{eq:sys_ol}, 
five constraints are imposed:

\begin{enumerate}
\item[(1)] \textbf{Phase margin:} $\pi + \angle G_{\mathrm{ol}}(j\omega_{\mathrm{c}}) \geq \Phi_{\mathrm{c}}$,
imposed at every frequency at which $|G_\mathrm{ol}|=1$.
\item[(2)] \textbf{Gain margin:} $20\log_{10}|G_{\mathrm{ol}}^{-1}(j\omega_{\phi})| \geq GM$
(e.g., GM=3~dB).
\item[(3)] \textbf{Resonant-band margin:} $\pi + \angle G_{\mathrm{ol}}(j\omega) \geq \Phi_{\mathrm{r}}$
for $0.9\omega_n < \omega < 1.1\omega_n$.
\item[(4)] \textbf{Harmonic retention:}
$\min\{|L(j\omega_{n-})|,|L(j\omega_{n+})|\}\geq\eta\,|L(j\omega_n)|$. (e.g., $\eta$=0.6)
\item[(5)] \textbf{Location of resonant frequency:}
$\omega_n=\arg\max_{\omega\in[0.9\omega_n,\,1.1\omega_n]}|G_{\mathrm{ol}}(j\omega)|$, 
i.e., the maximum must be located at $\omega_n$. 
\end{enumerate}

\subsection{Design of Initialization Strategy}
Two distinct initialization strategies are employed.
\textit{Random Initialization} uniformly samples within predefined bounds, 
offering simplicity but suffering from slow convergence due to the lack of domain knowledge.
Conversely, \textit{Experience-Based Initialization} integrates heuristics with model-driven accuracy.
Specifically, it leverages established principles (e.g., Benchmark \cite{pan_tpel_2017}, or EFDAM) to 
define a knowledge-based starting point.

The two strategies differ in more than convergence speed, and the difference is a direct consequence of
Theorem~\ref{thm:coupling_non_unique}. Because the search space contains near-flat directions along which
$J$ is almost constant, a randomly initialized local solver can terminate anywhere along such a direction.

\subsection{Design of Optimization Algorithms}
\subsubsection{Optimization Algorithm Selection}
To solve ICASM efficiently, two algorithms are proposed:
\begin{itemize}
    \item \textbf{MS-COBYLA:} 
    An improved deterministic solver enhancing sample efficiency. 
    By initializing via Latin Hypercube Sampling and executing parallel refinements, 
    it reduces COBYLA's \cite{ragonneau_mpc_2024} local-minima susceptibility
    and finds better optima more reliably than a single run; as a derivative-free,
    trust-region method it cannot certify global optimality.

    \item \textbf{PSO-COBYLA Hybrid:} 
    A sequential architecture balancing exploration and exploitation. 
    PSO \cite{cai_tpel_2021} performs stochastic global search to 
    provide a high-quality seed for COBYLA's deterministic, constraint-aware refinement. 
    This leverages PSO's exploratory power and COBYLA's native handling of hard constraints.
\end{itemize}

\subsubsection{Convergence and Complexity Analysis of the Algorithm}
\paragraph{Optimization Setup}
The multi-objective cost function is constructed with 
equal weighting factors $\lambda_i=1/3$ for the three normalized performance metrics. 
The optimization is subject to core constraints: phase margin $\text{PM}\geq30^\circ$, 
gain margin $\text{GM}\geq3$ dB, 
resonant band margin $\text{Res. PM}\geq10^\circ$, 
low-/high-order harmonic magnitude retention ratio $\eta=0.6$.

The normalization constants $\omega_{c\_\max}$, $L_{n\_\max}$ and $L_{1\_\max}$ in \eqref{eq:unified_obj}
are obtained before the search by three single-objective pre-scans, each maximizing one metric subject
only to constraints (1)--(2); they are therefore the same for all configurations and do not bias the
comparison. Constraint violations are handled in two stages: COBYLA treats the $\mathcal{O}(10^{2})$
inequalities natively (a trial point that violates any of them is simply not accepted), while the PSO
stage discards infeasible particles by re-initializing them inside the feasible set found so far, so the
swarm never reports a spurious high $J$ from an infeasible region.
The per-run budget is $200$ PSO iterations followed by up to a few hundred COBYLA iterations, with a
relative tolerance on both the objective and the trust-region radius as stopping criteria;
each configuration is repeated over $10$ independent runs, each launching $10$ refinement trajectories
from the Latin-hypercube (or EFDAM) initial set, and the best trajectory is retained. Reported values are
mean $\pm$ standard deviation over the $10$ runs.

\begin{figure}[htbp] % h,t,b: here, top, bottom
  \centering
  \includegraphics[width=1.0\linewidth]{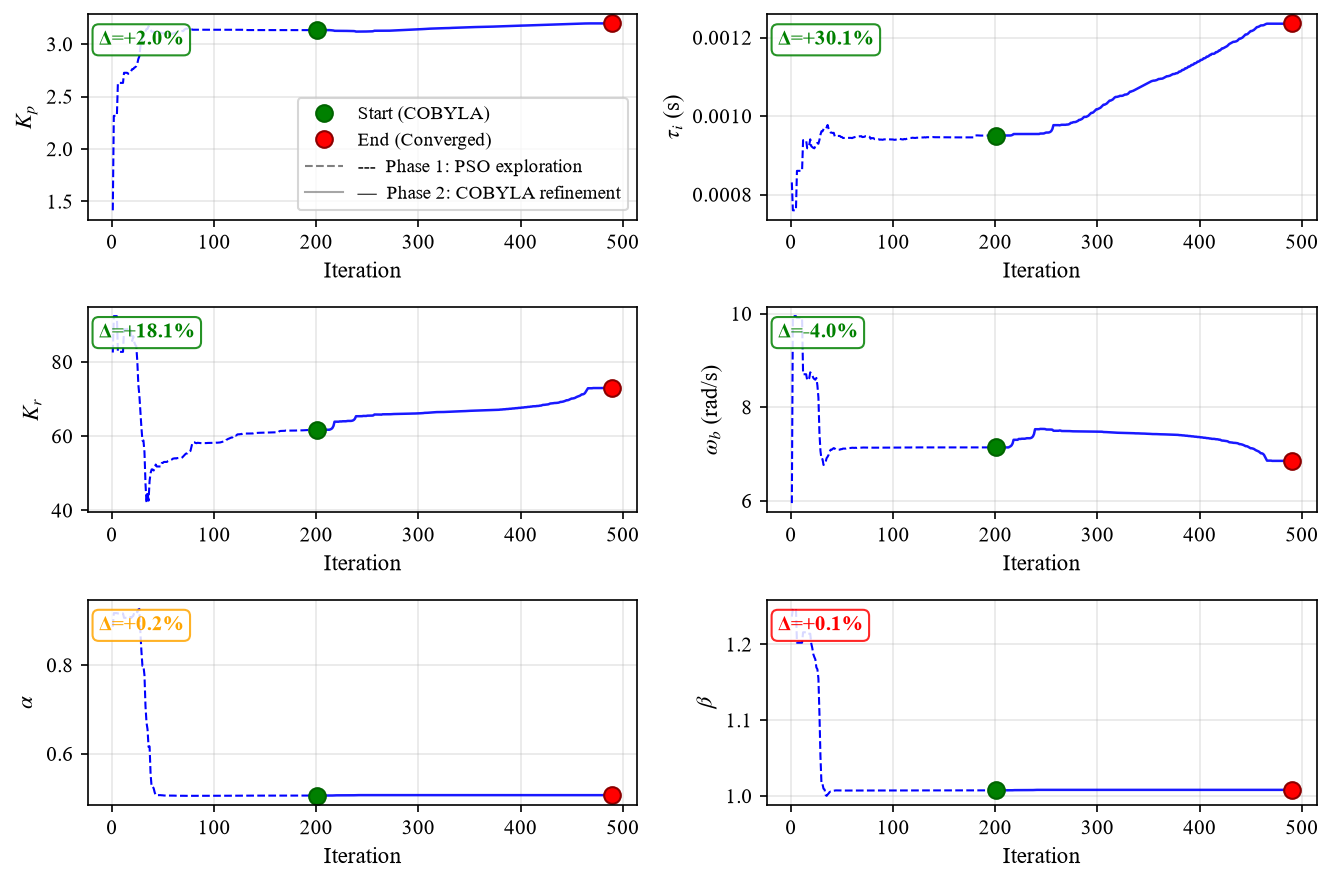}
  \caption{Parameter convergence history of the best trajectory.}
  \label{fig:ICASM_param_history}
\end{figure}

\begin{figure}[htbp] % h,t,b: here, top, bottom
  \centering
  \includegraphics[width=1.0\linewidth]{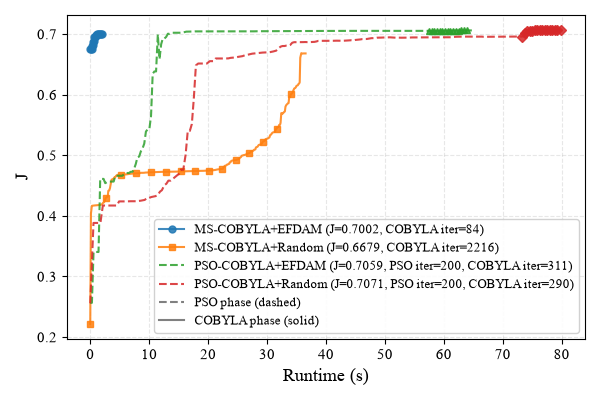}
  \caption{Comparison of the optimization trajectories.}
  \label{fig:ICASM_traj_comp}
\end{figure}

\paragraph{Convergence Performance Comparison}
Fig.~\ref{fig:ICASM_param_history} illustrates the evolution of the optimal parameters 
obtained from 10 independent PSO--COBYLA+Random runs. 
The experimental protocol is as follows: for each of the $4$ optimizer$\times$initialization
configurations, $10$ independent runs are executed with different random seeds; within each run,
$10$ refinement trajectories are launched in parallel from the initial set and only the trajectory
with the highest final $J$ is retained. Statistics in Table~\ref{tab:aggregated_results} are therefore
computed over the $10$ retained trajectories of the $10$ runs, whereas the single curves shown in
Figs.~\ref{fig:ICASM_param_history} and \ref{fig:ICASM_traj_comp} correspond to one representative
(best) trajectory. 
Consequently, the runtimes quoted for a single trajectory
(Fig.~\ref{fig:ICASM_traj_comp}) are necessarily smaller than the per-configuration means of
Table~\ref{tab:aggregated_results}, which include the full ten-run, ten-trajectory campaign.
The annotation $\Delta$ denotes the normalized adjustment magnitude during the COBYLA refinement stage, 
illustrating the fine-tuning process.

Fig.~\ref{fig:ICASM_traj_comp} compares the optimization trajectories of four optimizer-initialization strategies.
The horizontal axis denotes cumulative runtime; the vertical axis shows the objective $J$ (to be maximized).
Dashed and solid lines represent the PSO exploration and COBYLA refinement phases, respectively.

All values quoted in this paragraph refer to the single representative trajectory shown in
Fig.~\ref{fig:ICASM_traj_comp}; the corresponding statistics over the full ten-run campaign are given in
Table~\ref{tab:aggregated_results}.

PSO-COBYLA + Random achieves the best final performance ($J = 0.7071$, mean $0.7052\pm0.0016$)
in 81\,s for this trajectory (200 PSO + 290 COBYLA iterations).
PSO-COBYLA + EFDAM reaches a comparable $J = 0.7059$ (mean $0.7029\pm0.0024$)
in 65.2\,s (200 PSO + 311 COBYLA iterations),
confirming that PSO exploration mitigates sensitivity to poor initialization.
Conversely, MS-COBYLA + Random stagnates at $J = 0.6483\pm0.0090$ (mean over the ten runs)
after 2216 COBYLA iterations (36.6\,s for this trajectory),
indicating premature convergence to poor local minima.
In contrast, MS-COBYLA + EFDAM converges rapidly to $J = 0.7001\pm0.0000$ within
2.2\,s for this trajectory (mean $292.3\pm140.5$~s for the complete ten-run,
ten-trajectory campaign),
demonstrating high efficiency by applying a high-quality initialization.

These results highlight the exploration--exploitation trade-off: 
hybrid PSO-COBYLA provides robust global search but incurs higher computational costs, 
whereas MS-COBYLA achieves remarkable efficiency when provided with a high-quality initialization 
(e.g., EFDAM consistently outperforms random starts).

\begin{figure}[htbp] % h,t,b: here, top, bottom
  \centering
  \includegraphics[width=1.0\linewidth]{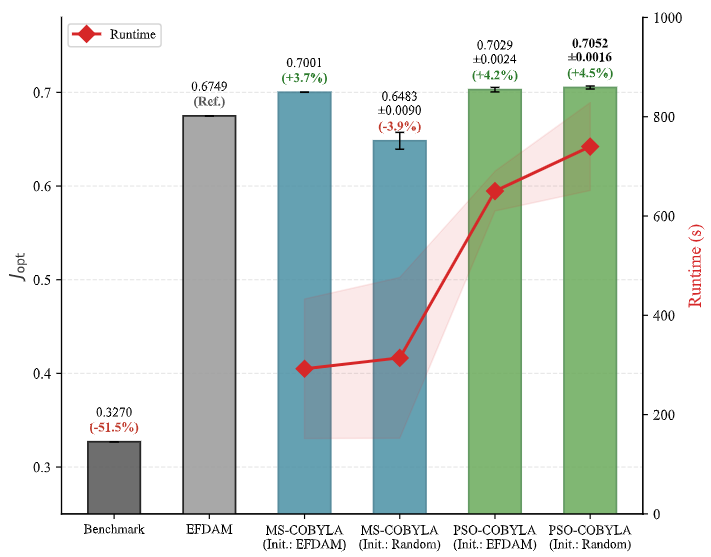}
  \caption{Performance comparison of different methods.}
  \label{fig:opt_comp}
\end{figure}

Fig.~\ref{fig:opt_comp} and Table~\ref{tab:aggregated_results} collectively illustrate the core performance metrics of the four optimizer-initialization strategies.
Sensitive parameters $\tau_\mathrm{i}$ and $K_\mathrm{r}$ exhibit excessively large standard deviations under random initialization, 
as exemplified by the MS-COBYLA + Random strategy, where $\tau_\mathrm{i}$ fluctuates by up to 1.773\,ms and 
$K_\mathrm{r}$ by 109.54.
Consequently, this configuration suffers a 3.9\% performance drop relative to the EFDAM reference ($J=0.6749$, Table~\ref{tab:aggregated_results}), 
consistent with the elevated parameter uncertainty.
While the MS-COBYLA + EFDAM strategy achieves the shortest average runtime (292.3 $\pm$ 140.5\,s) among all configurations, 
its optimal performance $J_{\text{opt}} = 0.7001 \pm 0.0000$ remains inferior to that of PSO-COBYLA-based variants.
Notably, this setup yields fully deterministic optimized parameters with zero variance across all parameters, 
demonstrating perfect repeatability albeit with suboptimal performance.

The PSO-COBYLA optimizer with random initialization (bold in Table~\ref{tab:aggregated_results}) is validated as the best strategy, 
achieving a peak objective value of $\mathbf{0.7052 \pm 0.0016}$.
This represents a 4.5\% improvement over EFDAM.
While sensitive parameters exhibit moderate fluctuations, 
they remain bounded and stable compared to the divergence of the MS-COBYLA + Random strategy.
Conversely, EFDAM initialization ensures near-deterministic convergence ($\tau_\mathrm{i}$ std.=0.135\,ms) at 
a slight peak-performance cost.
Although computationally intensive (739.9$\pm$88.4\,s), the runtime is acceptable for offline search.
Thus, random-initialized PSO-COBYLA is designated the preferred ICASM solution.
For context, Benchmark tuning via method \cite{pan_tpel_2017} yields $J=0.3270$, 
a 51.5\% deficit versus EFDAM, underscoring the limitations of manual discrete PIR design.

\begin{table}[htbp]
\centering
\caption{Performance of Optimizer $\times$ Initialization Strategies (Mean $\pm$ Std.)}
\label{tab:aggregated_results}
\setlength{\tabcolsep}{3pt}% 
\begin{tabular}{|>{\raggedright\arraybackslash}p{1.9cm}| l| c| c| c|}
\hline
Optimizer & Init. & $J_{\text{opt}}$ & Imp.$^\dagger$ (\%) & Runtime (s) \\
\hline
PSO-COBYLA & EFDAM  & 0.7029 $\pm$ 0.0024 & 4.2  & 650.2 $\pm$ 40.1 \\ 
PSO-COBYLA & Random & \textbf{0.7052 $\pm$ 0.0016} & 4.5  & 739.9 $\pm$ 88.4 \\
MS-COBYLA  & EFDAM  & 0.7001 $\pm$ 0.0000 & 3.7  & 292.3 $\pm$ 140.5 \\
MS-COBYLA  & Random & 0.6483 $\pm$ 0.0090 & -3.9 & 314.2 $\pm$ 162.0 \\
EFDAM       & ---    & 0.6749  & Ref. & --- \\
Benchmark \cite{pan_tpel_2017}      & ---    & 0.3270  & -51.5& --- \\
\hline
\multicolumn{5}{@{}l@{}}{$^\dagger$ Imp.: improvement relative to Ref.}
\end{tabular}
\end{table}

\begin{table*}[htbp]
\centering
\caption{Optimized ICASM Controller Parameters by Optimizer $\times$ Initialization (Mean $\pm$ Std.)}
\label{tab:ICASM_params}
\begin{tabular}{|l|l|c|c|c|c|c|c|}
\hline
Optimizer & Initialization& $K_\mathrm{p}$    & $\tau_\mathrm{i}$ (ms)       & $K_\mathrm{r}$      & $\omega_\mathrm{b}$ (rad/s) & $\alpha$ & $\beta$ \\
\hline
PSO-COBYLA & EFDAM       & 3.119 $\pm$ 0.039 & 0.971 $\pm$ 0.135   & 25.35 $\pm$ 0.70    & 19.56 $\pm$ 0.28 & 0.5000 $\pm$ 0.0000 & 1.0080 $\pm$ 0.0001 \\
PSO-COBYLA & Random      & 3.169 $\pm$ 0.032 & 1.126 $\pm$ 0.123   & 26.65 $\pm$ 1.02    & 18.68 $\pm$ 0.68 & 0.5002 $\pm$ 0.0003 & 1.0080 $\pm$ 0.0000 \\
MS-COBYLA  & EFDAM       & 3.060 $\pm$ 0.000 & 0.817 $\pm$ 0.000   & 24.21 $\pm$ 0.00    & 20.00 $\pm$ 0.00 & 0.5000 $\pm$ 0.0000 & 1.0080 $\pm$ 0.0000 \\
MS-COBYLA  & Random      & 2.810 $\pm$ 0.041 & 0.814 $\pm$ 0.042   & 60.55 $\pm$ 52.80   & 26.24 $\pm$ 12.56& 0.5138 $\pm$ 0.0070 & 1.0083 $\pm$ 0.0001 \\
EFDAM       & ---         & 3.022             & 0.811               & 22.36               & 19.96            & 0.5000              & 1.0075              \\
Benchmark \cite{pan_tpel_2017}     & ---         & 2.500             & 2.122               & 15.00               & 12.57            & 0.5000              & 1.0000              \\
\hline
\multicolumn{8}{@{}l@{}}{\footnotesize Fixed benchmark/manual tuning values are reported without standard deviation.} \\
\end{tabular}
\end{table*}

\paragraph{Algorithmic Superiority Analysis}
Comparative results reveal a distinct performance hierarchy among the evaluated strategies. 
The proposed PSO-COBYLA framework outperforms MS-COBYLA, 
leveraging PSO's global exploration capability to escape the local minima inherent in the non-convex parameter space. 
Consequently, the PSO-COBYLA configuration is adopted as the
definitive ICASM synthesis methodology, which delivers enhanced precision
while maintaining favorable computational tractability.

%-------------------------------------------------------------------------------
% Section V: Experimental Validation
%-------------------------------------------------------------------------------
\section{Experimental Evaluation}\label{sec5}
\subsection{Testbed Configuration}
\begin{figure}[htbp] % h,t,b: here, top, bottom
  \centering
  \includegraphics[width=1.0\linewidth]{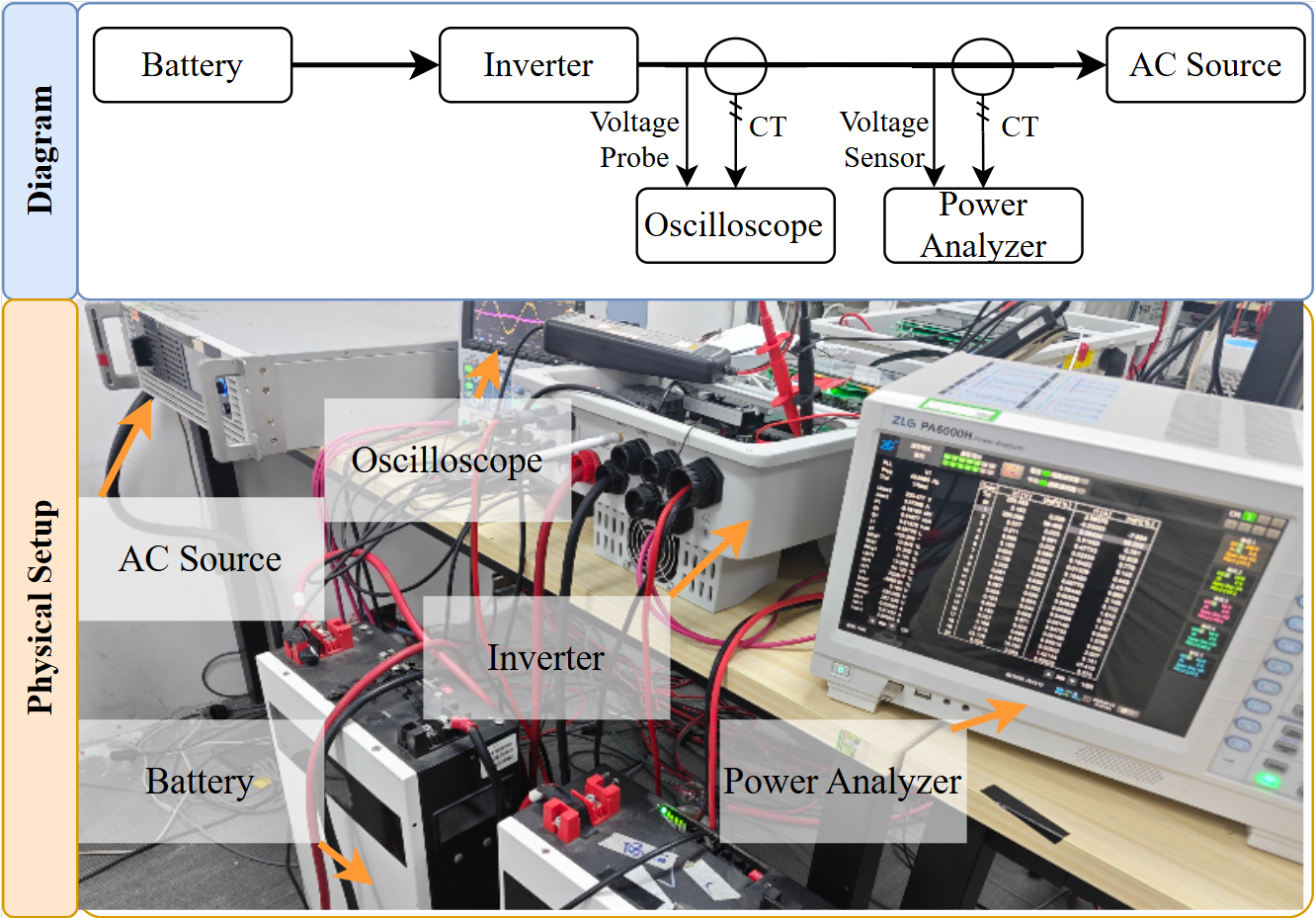}
  \caption{Diagram and physical setup of the inverter testbed.}
  \label{fig:inv_test_setup}
\end{figure}

To validate the efficacy of the proposed EFDAM and ICASM methods, 
a 5-kW single-phase grid-tied inverter testbed is constructed, 
as illustrated in Fig.~\ref{fig:inv_test_setup}. 
The prototype integrates a DC/DC boost pre-regulator and a DC/AC inversion stage, 
with the DC bus powered by a lithium-ion battery pack. 
A grid simulator (model: Chroma 61815) emulates the utility grid under both steady-state and transient operating conditions. 
Waveform acquisition is handled by a Yokogawa DLM3024 oscilloscope, 
while a ZLG PA6000H power analyzer captures total harmonic distortion (THD).

\subsection{Design Results Comparison}
Table~\ref{tab:result_comp} benchmarks the proposed ICASM optimization results against 
the analytical EFDAM approximation and conventional Benchmark-tuned parameters.
%\textbf{Construction of the Benchmark, and what each comparison isolates.}
The Benchmark is obtained by applying the unified design procedure of
\cite{pan_tpel_2017} to the same plant model \eqref{eq:cont_filter}, the same sampling
period $T_\mathrm{s}$, the same computational delay, the same evaluation grid and the same
stability-margin requirements as the proposed methods; its discretization is the Tustin rule
($\alpha=0.5$) without prewarping ($\beta=1$), which is a common industrial choice.
More importantly, the EFDAM row should be read as follows: as shown in Section~\ref{sec3},
EFDAM yields $\alpha_\mathrm{opt}=0.5$ and $\beta_\mathrm{opt}$ given by \eqref{eq:beta_opt}, i.e., the
classical prewarped-Tustin discretization applied to gains that were themselves tuned with the
digital delay budget accounted for. EFDAM is thus precisely the ``best classical two-step design'':
continuous-domain loop shaping followed by a correctly prewarped standard discretization.
Consequently, the improvement of ICASM over EFDAM ($J$: $0.6749\rightarrow0.7052$, $+4.5\%$;
THDi at the $19$th harmonic: $7.76\%\rightarrow7.21\%$) measures \emph{exactly} the benefit of the joint
dual-domain optimization, with the effect of the discretization rule already factored out. The much larger
improvement of EFDAM over the Benchmark ($J$: $0.3270\rightarrow0.6749$) then measures the combined effect
of discretization-rule correction and delay-aware loop shaping.

The Benchmark design adopts an excessively large integral time constant $\tau_\mathrm{i}=2.122$ ms to guarantee unconditional stability, 
which severely degrades low-frequency reference tracking and disturbance rejection performance. 
Conversely, EFDAM and ICASM reduce $\tau_\mathrm{i}$ to 0.811--1.288\,ms while satisfying all stability and harmonic constraints, 
overcoming the inherent conservatism of the Benchmark design.
Relative to the best-found ICASM optimum, the EFDAM solution exhibits moderate deviations across the parameter set. 
Notably, the strongly coupled resonant parameters $(K_\mathrm{r}, \omega_\mathrm{b})$ show discrepancies of -17.0\% and +7.8\%, respectively. 
This is a direct consequence of the analytical boundary approximation, 
which prioritizes feasibility over absolute optimality for these highly interactive terms. 
Critically, the deviations in the PI controller parameters ($K_\mathrm{p}$: -5.8\%, $\tau_\mathrm{i}$: -37.0\%) are 
partially offset by the near-perfect fidelity of the digital discretization parameters ($\alpha$: 0.0\%, $\beta$: -0.1\%). 
This validates the high fidelity of the EFDAM method: despite its low-complexity nature, 
it delivers a near-optimal feasible solution that accurately captures the dominant control dynamics.

\begin{table}[htbp]
\centering
\caption{Controller Parameter Design Comparison}
\label{tab:result_comp}
\begin{tabular}{|l|c|c|c|c|}
\hline
Parameter                  & Benchmark \cite{pan_tpel_2017}  & EFDAM   & ICASM   & Err. (\%) \\
\hline
$K_\mathrm{p}$             & 2.500    & 3.022   & 3.209   & -5.8   \\
$\tau_\mathrm{i}$ (ms)     & 2.122    & 0.811   & 1.288   & -37.0  \\
$K_\mathrm{r}$             & 15.00    & 22.36   & 26.93   & -17.0  \\
$\omega_\mathrm{b}$ (rad/s)& 12.57    & 19.96   & 18.51   & 7.8    \\
$\alpha$                   & 0.500    & 0.500   & 0.500   & 0.0    \\
$\beta$                    & 1.000    & 1.0075  & 1.0080  & -0.1   \\
\hline
\multicolumn{5}{@{}l@{}}{\footnotesize Err.: denotes relative error of EFDAM vs. ICASM optimum.} \\
\end{tabular}
\end{table}

\subsection{Steady-State Performance}
\begin{figure}[htbp]
  \centering
  % line 1
  \begin{minipage}[t]{1.0\linewidth}
    \centering
    \includegraphics[width=\linewidth]{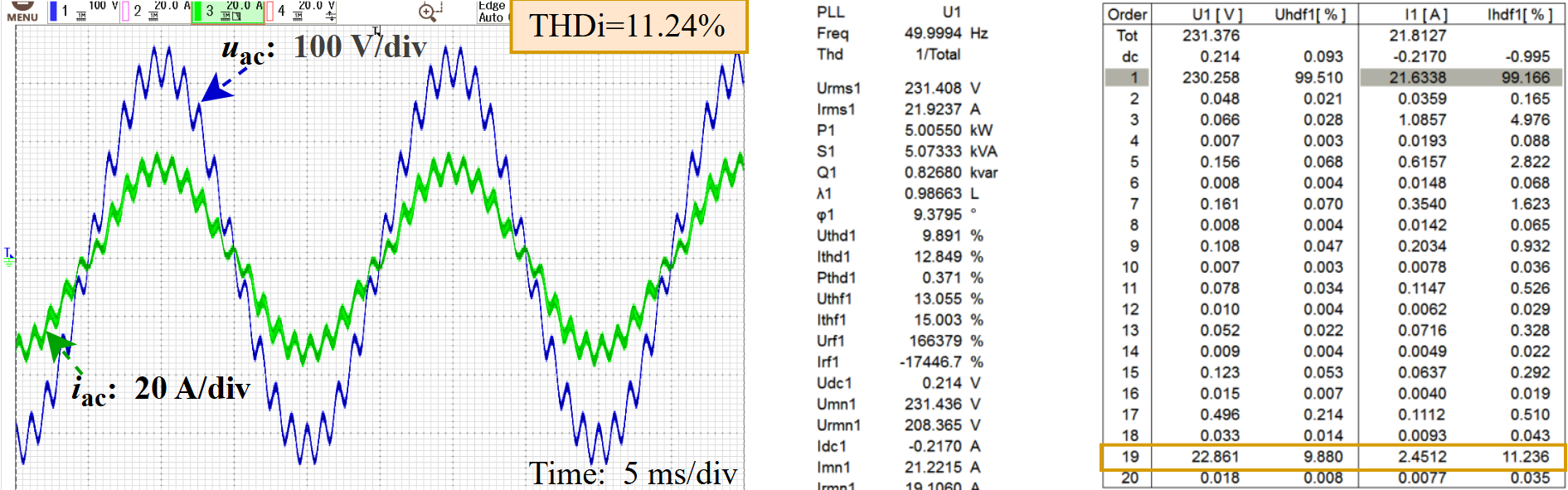}
    \centerline{\footnotesize (a) Benchmark.}
  \end{minipage}
  \hfill
  % line 2
  \begin{minipage}[t]{1.0\linewidth}
    \centering
    \includegraphics[width=\linewidth]{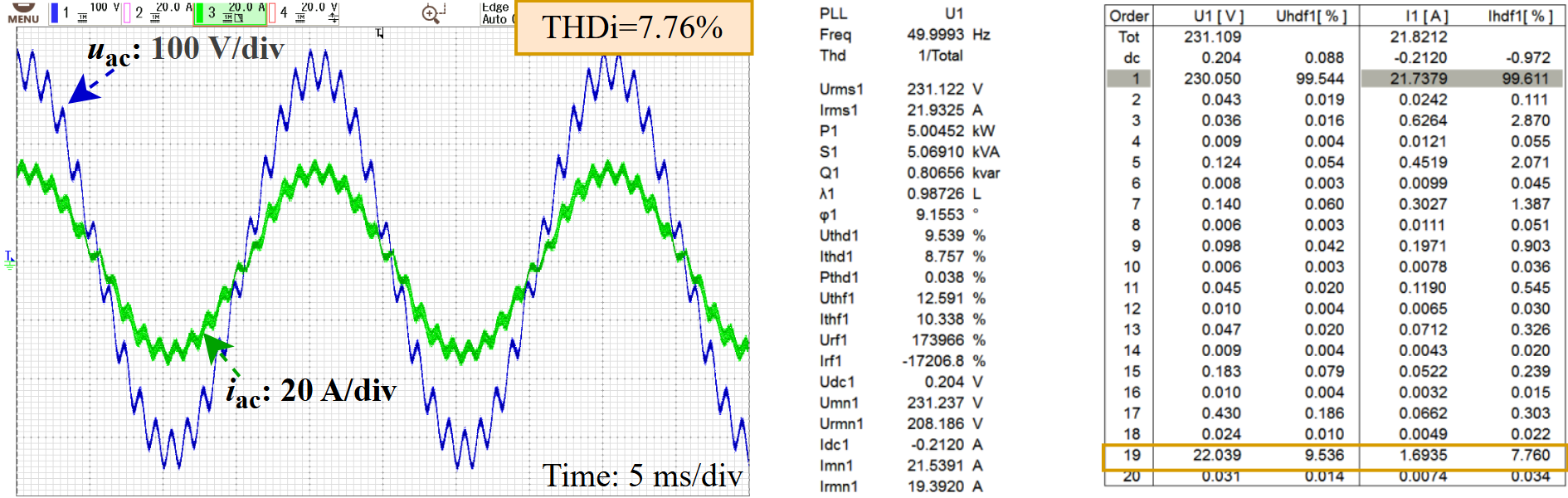}
    \centerline{\footnotesize (b) EFDAM.}
  \end{minipage}
  % line 3
  \begin{minipage}[t]{1.0\linewidth}
    \centering
    \includegraphics[width=\linewidth]{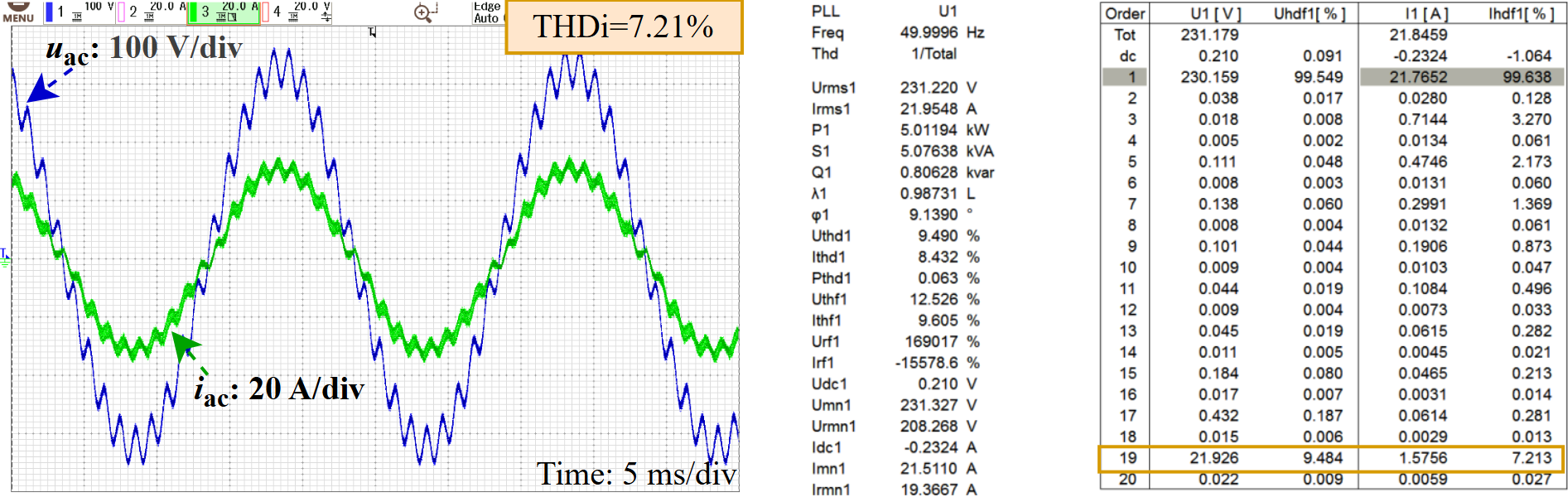}
    \centerline{\footnotesize (c) ICASM.}
  \end{minipage}
  %\vspace{0.5em}
  \caption{Experimental waveforms (left panel) and measured data (right panel) of AC voltage and current under different methods.}  
  \label{fig:inv_test_waveform_n19}
\end{figure}

The steady-state performance in terms of current waveforms and THDi is compared in Fig.~\ref{fig:inv_test_waveform_n19}.
Subfigures~\ref{fig:inv_test_waveform_n19}(a)--(c) correspond to Benchmark, EFDAM, and ICASM, respectively.
In each subfigure, the left panel displays the experimental waveform, 
while the right panel shows the data captured by the power analyzer.
Visually, distinct differences exist between the Benchmark method and the other two (EFDAM and ICASM).
However, the waveforms of EFDAM and ICASM are nearly identical;
conversely, the power analyzer data reveal subtle distinctions between them.
Under rated power ($5\,\mathrm{kW}$) and comparable grid voltage distortion,
the THDi is reduced from $11.24\,\%$ (Benchmark) to $7.76\,\%$ (EFDAM) and $7.21\,\%$ (ICASM).
The calculated improvement rates are \textbf{31.0\,\%} for EFDAM and \textbf{35.9\,\%} for ICASM relative to the Benchmark baseline,
demonstrating the superior harmonic suppression capability of the proposed optimal approaches.

\subsection{Dynamic Response Performance}
\begin{figure}[htbp]
  \centering
  % line 1
  \begin{minipage}[t]{1.0\linewidth}
    \centering
    \includegraphics[width=\linewidth]{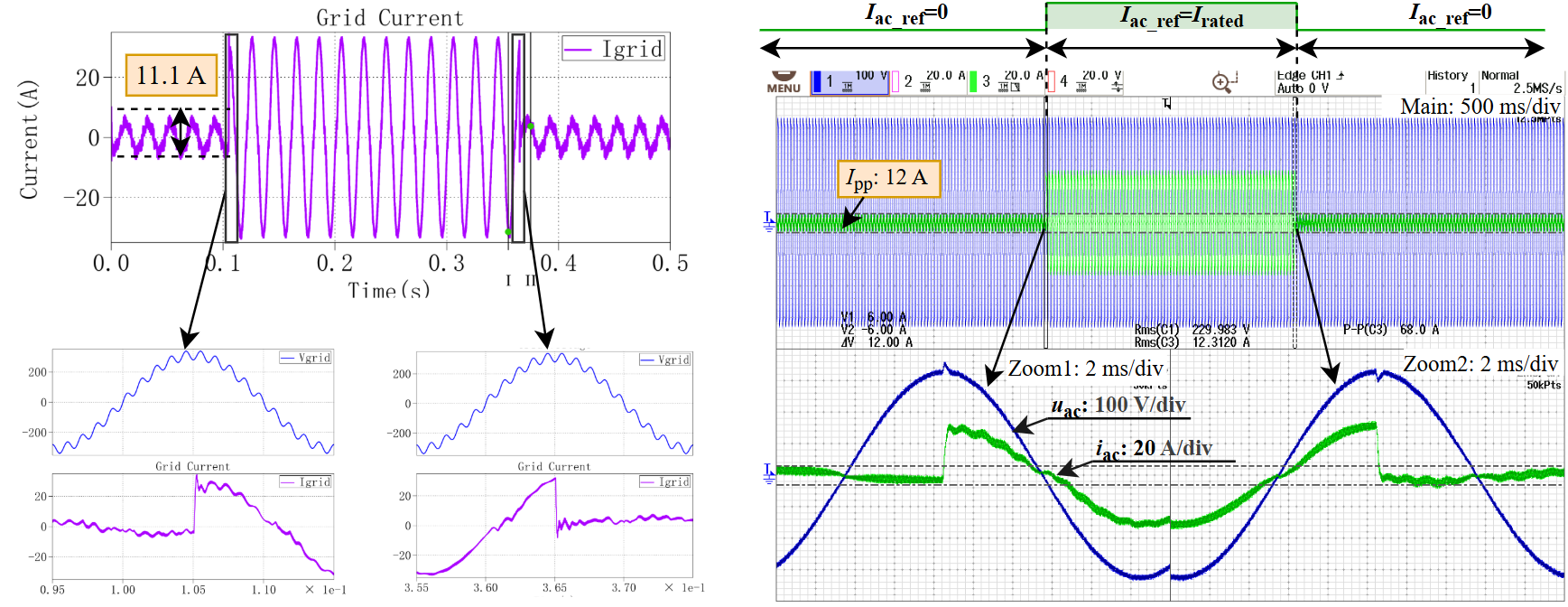}
    \centerline{\footnotesize (a) Benchmark.}
  \end{minipage}
  \hfill
  % line 2
  \begin{minipage}[t]{1.0\linewidth}
    \centering
    \includegraphics[width=\linewidth]{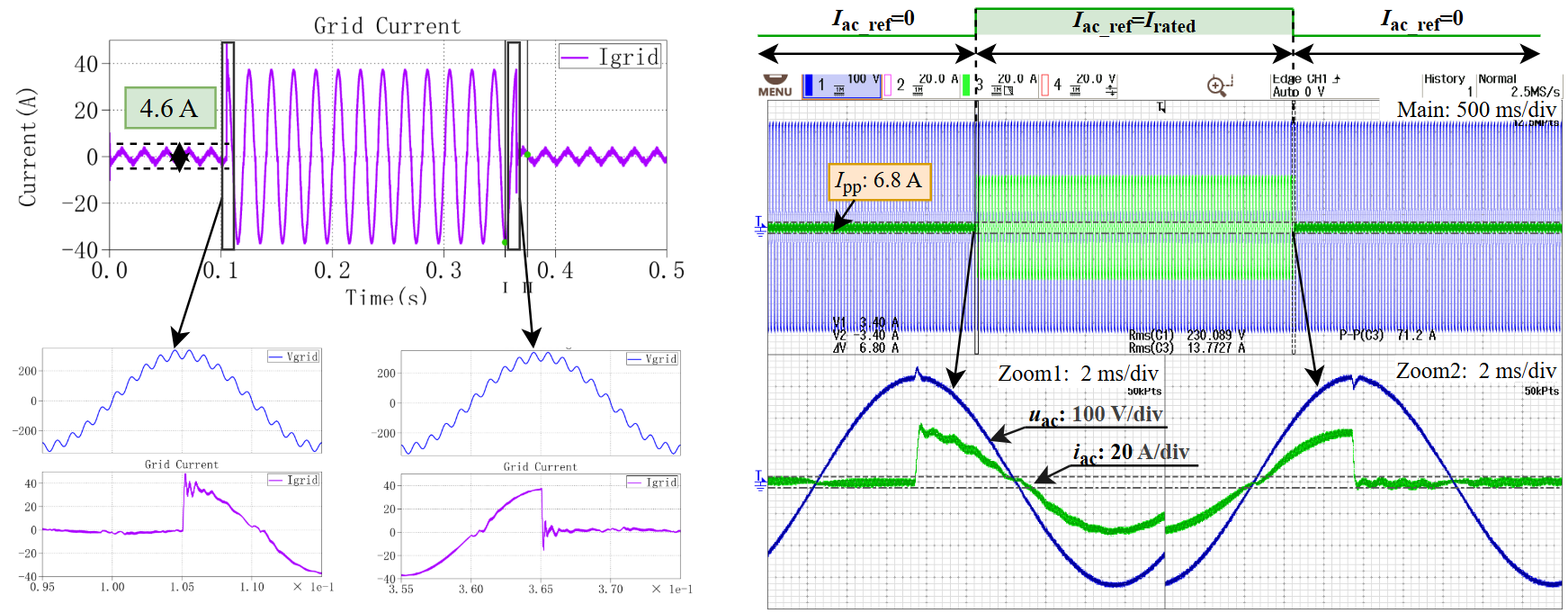}
    \centerline{\footnotesize (b) EFDAM.}
  \end{minipage}
  %\vspace{0.5em}
  \caption{Simulation (left panel) and experimental (right panel) comparison of dynamic command tracking under different methods.}
  \label{fig:dynamic_test_current_track}
\end{figure}

The dynamic command tracking performance is evaluated under a step change scenario 
($0 \rightarrow I_\mathrm{rated} \rightarrow 0$).
As depicted in Fig.~\ref{fig:dynamic_test_current_track}, 
the simulation and experimental waveforms integrate the global transient process with two zoomed-in views: 
the rising edge ($0 \rightarrow I_\mathrm{rated}$) and the falling edge ($I_\mathrm{rated} \rightarrow 0$).
Given the marginal visual distinction between EFDAM and ICASM in steady-state analysis, 
only the representative EFDAM results are presented here for brevity.

Quantitatively, the peak-to-peak current ripple $I_\mathrm{pp}$ at 
zero active power command ($I_{\mathrm{d\_ref}}=0$) is reduced from 
12~A for the Benchmark method [Fig.~\ref{fig:dynamic_test_current_track}(a)] to 
6.8~A for EFDAM [Fig.~\ref{fig:dynamic_test_current_track}(b)], corresponding to a 43.3\% improvement.
These results demonstrate the superior ripple-suppression capability of the proposed EFDAM 
(and by extension, ICASM) over the Benchmark method at the zero-active-power operating point.

\begin{figure}[htbp] % h,t,b: here, top, bottom
  \centering
  \includegraphics[width=1.0\linewidth]{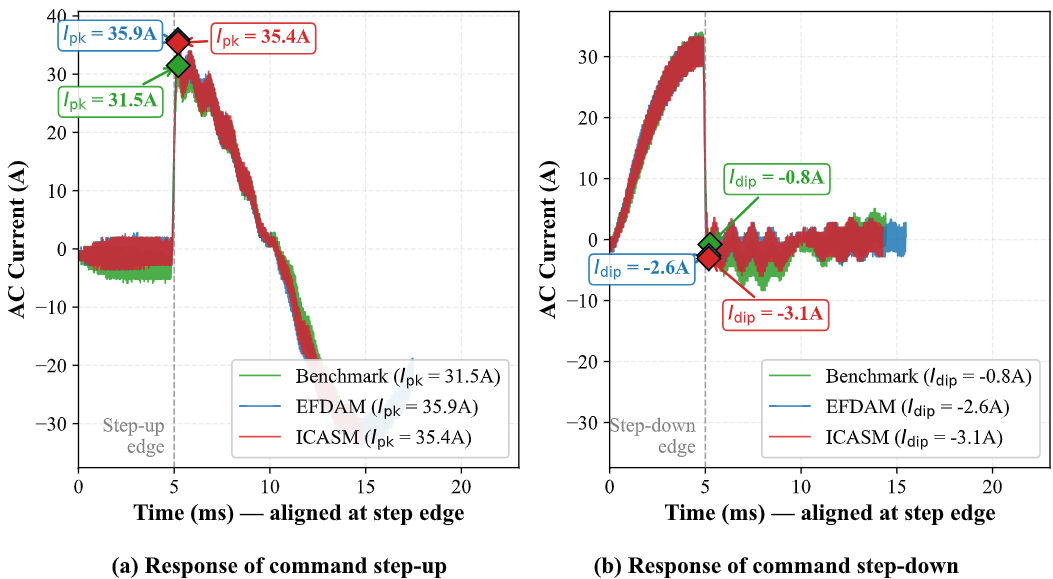}
  \caption{Comparison of AC current waveforms using different methods during current command transients.}
  \label{fig:dynamic_test_cmd_step_comp} 
\end{figure}

Fig.~\ref{fig:dynamic_test_cmd_step_comp} compares the AC current waveforms during step changes in the current command.
Fig.~\ref{fig:dynamic_test_cmd_step_comp}(a) illustrates the step-up transient. 
While the Benchmark method exhibits the lowest peak current ($I_{\mathrm{pk}} = 31.5\,\mathrm{A}$), 
the EFDAM and ICASM result in slightly higher peaks, ranging from 35.4~A to 35.9~A.
Fig.~\ref{fig:dynamic_test_cmd_step_comp}(b) depicts the step-down transient. 
In this case, the Benchmark method yields a minimum current of $-0.8\,\mathrm{A}$, 
whereas EFDAM and ICASM exhibit a more pronounced undershoot, 
ranging from $-3.1\,\mathrm{A}$ to $-2.6\,\mathrm{A}$.

These results indicate that although the proposed EFDAM and ICASM introduce slightly higher overshoot/undershoot 
compared to the Benchmark, their dynamic characteristics are virtually identical, 
confirming the consistency of the proposed optimization strategies.

\begin{figure}[htbp] % h,t,b: here, top, bottom
  \centering
  \includegraphics[width=0.8\linewidth]{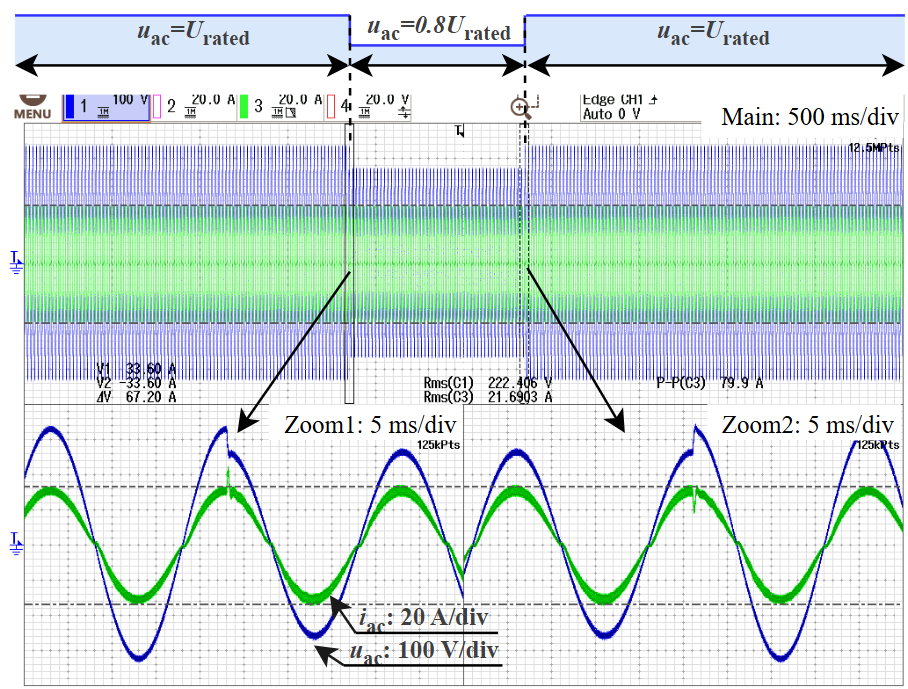}
  \caption{Experimental waveforms using EFDAM during grid voltage transients.}
  \label{fig:dynamic_test_voltage_step}
\end{figure}

\begin{figure}[htbp] % h,t,b: here, top, bottom
  \centering
  \includegraphics[width=1.0\linewidth]{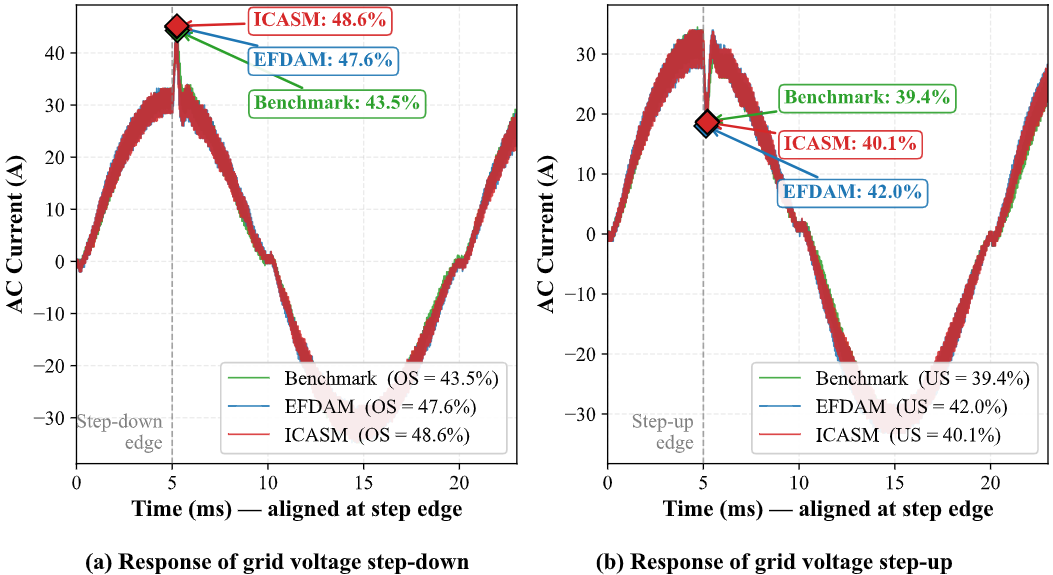}
  \caption{Comparison of AC current waveforms using different methods during grid voltage transients. OS: overshoot; US: undershoot.}
  \label{fig:dynamic_test_grid_step_comp} 
\end{figure}

The robustness against grid voltage disturbances is evaluated in Fig.~\ref{fig:dynamic_test_voltage_step}, 
where the EFDAM method maintains current stability under voltage fluctuations.
Fig.~\ref{fig:dynamic_test_grid_step_comp} further compares the transient metrics.

Quantitatively, the overshoot (OS) during the step-down transient increases modestly 
from 43.5\% (Benchmark) to 47.6\%--48.6\% for the proposed methods.
Similarly, the undershoot (US) during the step-up transient shifts from 39.4\% to 40.1\%--42.0\%.
This phenomenon suggests a typical trade-off: 
the enhanced steady-state harmonic suppression capability of EFDAM and ICASM is accompanied by a slight reduction in damping ratio, 
leading to marginally higher overshoot and undershoot. 
Nevertheless, the near-identical responses of EFDAM and ICASM underscore the reliability of the proposed optimization framework.

\subsection{Generality Validation Across Harmonic Configurations}
\begin{figure}[htbp] % h,t,b: here, top, bottom
  \centering
  \includegraphics[width=1.0\linewidth]{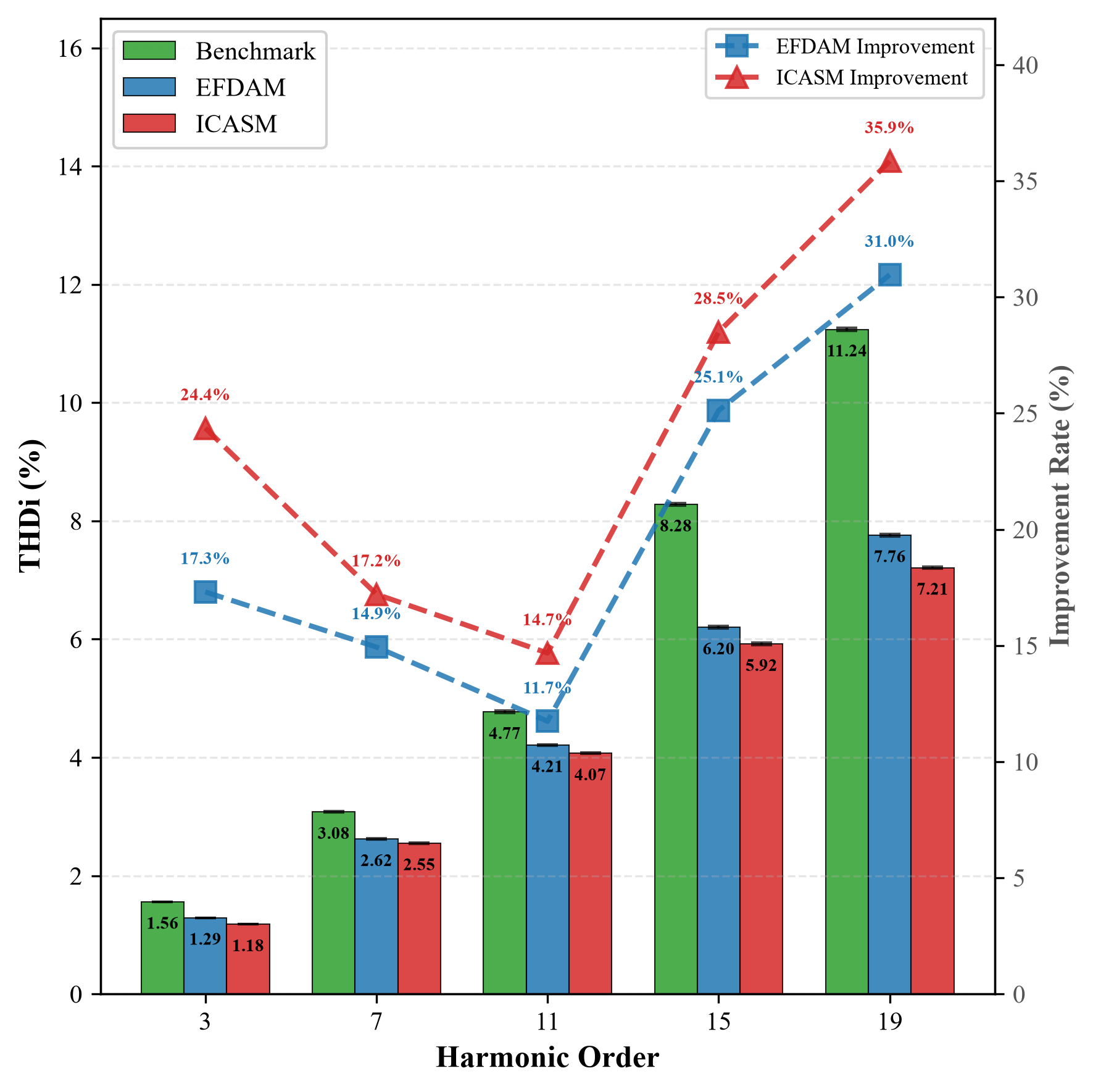}
  \caption{Generality validation across different harmonic configurations.}
  \label{fig:harmonic_comp} 
\end{figure}

To validate the generality of the proposed optimization framework, 
the THDi suppression performance is evaluated across a wide spectrum of characteristic harmonics 
(i.e., the 3rd, 7th, 11th, 15th, and 19th orders), as depicted in Fig.~\ref{fig:harmonic_comp}.
For each harmonic order $h$, the green, blue, and red bars correspond to the Benchmark baseline, EFDAM, and ICASM, respectively.
It is evident that both EFDAM and ICASM consistently yield significantly lower THDi values than the Benchmark method across all tested scenarios.
Furthermore, the dynamic trajectories and steady-state performance trends of EFDAM and ICASM exhibit remarkable consistency.
Quantitatively, the absolute performance deviation $\Delta\mathrm{THDi} = | \mathrm{THDi}_\mathrm{EFDAM} - \mathrm{THDi}_\mathrm{ICASM} |$ 
between the two proposed methods remains marginal, ranging from 0.07\% to 0.55\% 
(specifically, 0.18\% at $h=3$, 0.07\% at $h=7$, 0.14\% at $h=11$, 0.28\% at $h=15$, and 0.55\% at $h=19$).
This minimal inter-method discrepancy, even at high-order harmonics, 
not only validates the robustness of the optimization logic but also 
confirms the universal applicability of the proposed EFDAM and ICASM algorithms in optimizing control parameters of resonant controllers.

%-------------------------------------------------------------------------------
% Section : Conclusion
%-------------------------------------------------------------------------------
\section{Conclusion}\label{sec6}
This article proposes a dual-domain co-design based on the formalized non-uniqueness theorem, 
which establishes that discrete performance is governed by the coupling of 
continuous control parameters ($\Theta_\mathrm{c}$) and discretization parameters ($\Theta_\mathrm{d}$). 
Subsequently, we propose two methodologies: the EFDAM for rapid closed-form deployment and the ICASM for constrained black-box optimization. 
To efficiently satisfy the stability constraints, which are non-convex, 
we propose two algorithms---MS-COBYLA and PSO-COBYLA---that balance global exploration with local refinement.

Comprehensive validation on a 5-kW grid-tied inverter prototype validates the efficacy of the proposed framework. 
While the pursuit of higher loop gains inevitably introduces marginal transient overshoot during step changes, 
the system retains robust dynamic stability and damping characteristics. 
Notably, the THDi difference between the two methods remains below $0.55$ percentage points
across the tested harmonic orders (individual parameters differ by up to tens of percent, which is
consistent with the redundancy established by the theorem),
confirming the consistency between analytical approximations and numerical optimizations. 
By systematically removing the dependence of the design on trial-and-error heuristics,
this work establishes a systematic, reproducible, and high-performance guideline for 
optimizing digital control of power converters.
However, the present study is restricted to a linear time-invariant plant model with synchronous, loss-free
sampling and a single active resonant branch; robustness against parameter drift ($L_\mathrm{f}$, grid
impedance), measurement noise, packet loss and plant nonlinearities is not addressed here and requires the
formulation to be extended (e.g., by worst-case or stochastic constraints), which is left for future work.

\end{document}